\documentclass[jornal]{IEEEtran}
\usepackage{amsfonts}
\usepackage{amssymb}
\usepackage{amsmath}
\usepackage{mathrsfs}
\usepackage{verbatim}
\usepackage{subfigure}
\usepackage{balance}
\usepackage{booktabs}
\usepackage{fancybox}
\usepackage{bm}
\usepackage{extarrows}
\usepackage{algorithm}
\usepackage{algorithmic}
\usepackage{multirow}
\usepackage{array}
\usepackage{graphicx}
\usepackage{epstopdf}

\newtheorem{theorem}{Theorem}
\newtheorem{lemma}{Lemma}

\begin{document}
%\baselineskip 4.1ex
%\title{On the User Selection and Multiuser Gain of Secrecy Communication in Uplink Networks}
\title{Secure Communication in Uplink Transmissions: User Selection and Multiuser Secrecy Gain}
%On the role of helper in cooperative secrecy transmission with jamming
%
%
\author{Hao~Deng,
        Hui-Ming~Wang,~\IEEEmembership{Senior Member,~IEEE,}
        Jinhong~Yuan,~\IEEEmembership{Fellow,~IEEE,}\\
        Wenjie~Wang,~\IEEEmembership{Member,~IEEE,}
        and Qinye~Yin
%添加作者
%\author{Michael~Shell,~\IEEEmembership{Member,~IEEE,}
%        John~Doe,~\IEEEmembership{Fellow,~OSA,}
%        and~Jane~Doe,~\IEEEmembership{Life~Fellow,~IEEE}% <-this % stops a space

\thanks {Manuscript received January 11, 2016; revised May 8, 2016 and June 18, 2016; accepted June 18, 2016. Date of publication XXXXX, 2016; date of current version XXXXX, 2016. The work of H. Deng, H.-M. Wang and W. Wang was partially supported by the Foundation for the Author of National Excellent Doctoral Dissertation of China under Grant 201340, the New Century Excellent Talents Support Fund of China under Grant NCET-13-0458, the Fok Ying Tong Education Foundation under Grant 141063, the Fundamental Research Funds for the Central University under Grant 2013jdgz11, the National High-Tech Research and Development Program of China under Grant No.2015AA01A708, and the Young Talent Support Fund of Science and Technology of Shaanxi Province under Grant 2015KJXX-01. The work of J. Yuan was supported by Australia Research Council (ARC) Discovery Project DP160104566. The associate editor coordinating the review of this paper and approving it for publication was K. Tourki. \emph{(Corresponding author: Hui-Ming Wang)}}
\thanks {H. Deng is with the Ministry of Education Key Laboratory for Intelligent Networks and Network Security, School of Electronic and Information Engineering, Xi'an Jiaotong University, Xi'an 710049, China, and also with the School of Physics and Electronics, Henan University, Kaifeng 475001, China (e-mail: gavind@163.com).}
\thanks  {H.-M. Wang, W. Wang and Q. Yin are with the Ministry of Education Key Laboratory for Intelligent Networks and Network Security, School of Electronic and Information Engineering, Xi'an Jiaotong University, Xi'an 710049, China (e-mail: xjbswhm@gmail.com; wjwang@xjtu.edu.cn; qyeyin@gmail.com).}
\thanks {J. Yuan is with the School of Electrical Engineering and Telecommunications, University of New South Wales, Sydney, Australia (e-mail: j.yuan@unsw.edu.au).}
\thanks {Color versions of one or more of the figures in this paper are available online at http://ieeexplore.ieee.org.}
\thanks {Digital Object Identifier 10.1109/TCOMM.2016.XXXXX}
}

\maketitle

\begin{abstract}
%\boldmath
In this paper, we investigate secure communications in uplink transmissions, where there are a base station (BS) with $M$ receive antennas, $K$ mobile users each with a single antenna, and an eavesdropper with $N$ receive antennas. The closed-form expressions of the achievable ergodic secrecy sum-rates (ESSR) for a \emph{random} $k$ users selection scheme in the high and low SNR regimes are presented. It is shown that the scaling behavior of ESSR with respect to the number of served users $k$ can be quite different under different system configurations, determined by the numbers of the BS antennas and that of the eavesdropper antennas. In order to achieve a multiuser gain, two low-complexity user selection schemes are proposed under different assumptions on the eavesdropper's channel state information (CSI). The closed-form expressions of the achievable ESSRs and the multiuser secrecy gains of the two schemes are also presented in both low and high SNR regimes. We observe that, as $k$ increases, the multiuser secrecy gain increases, while the ESSR may decrease. Therefore, when $N$ is much larger than $M$, serving one user with the strongest channel (TDMA-like) is a favourable secrecy scheme, where the ESSR scales with $\sqrt{{2\log K}}$.
\end{abstract}
%Since the exact distribution of achievable secrecy rate in a Rayleigh fading environment is difficult to analyze, a closed-form expression of ESSR can not be obtained.

% \fi
\IEEEpeerreviewmaketitle

\vspace{8ex}
\section{Introduction}
\IEEEPARstart{W}IRELESS physical layer security has received considerable attention recently, especially in multiple-input multiple-output (MIMO) communication systems \cite{ref:F. Oggier}-\cite{ref:Wang TWC}. The basic idea of secure MIMO transmissions is to construct a better equivalent channel by utilizing the spatial degrees of freedom provided by multiple antennas, such that the quality of the main channel (from the transmitter to the legitimate receiver) is better than that of the wiretap channel (from the transmitter to the eavesdropper). Various signal processing approaches have been developed to enhance the security, such as transmit precoding (TR) \cite{ref:A. L.}-\cite{ref:Wu2016TIT}, and artificial noise (AN) \cite{ref:Wang Hybrid}-\cite{ref:Wang TWC}. In all the above literature, it is shown that the secrecy performance, in term of secrecy rate, or secrecy outage probability, can be improved via TR or AN schemes.

In cellular systems, multiple antennas can be easily deployed at the base station (BS). Thus in a downlink transmission, the BS can facilitate secure transmissions by using TR or AN. In contrast, the mobile users in most cases equip with a single antenna due to the limitations of size and cost. The lack of spatial degrees of freedom for mobile users makes it difficult to guarantee security in uplink transmissions. Especially when the eavesdropper has more antennas than the BS does, secure transmission is very difficult to be guaranteed. Under such a scenario, it is natural to exploit multiuser gain to improve the security. In particular, when there are multiple users in an uplink transmission, \emph{multiuser secrecy gain} can be achieved by serving users with better channels. However, few of existing works have addressed the issues of user selection and multiuser gain in secure communications, which motivates our studies.

For a $K$-user uplink transmission without secrecy consideration, the multiuser gain has been extensively studied. It is known that the sum-rate scales like $\log \log K$ with user selection/scheduling \cite{ref:A. Goldsmith}-\cite{ref:M. Sharif Partial}. However, this result \emph{can not} be generalized directly to secure communications, because secrecy capacity is defined as the maximum rate difference between the achievable transmission rates of the main channel and that of the wiretap channel. Hence, it is still not clear how many users should be scheduled concurrently and what the exact scaling law of multiuser gain for secure transmission is. Therefore, this paper considers an uplink transmission where multiple users simultaneously send confidential messages to a multi-antenna BS in the presence of a multi-antenna eavesdropper. More specifically, we analyze the secrecy sum-rate of the uplink transmission and demonstrate the scaling laws of the multiuser secrecy gains with respect to the number of served users, $k$, and the total number of users, $K$.

\subsection{Related works}
%In a multiuser system, it is well known that multiuser diversity can be used to improve the system sum-rate by user selection/scheduling, when the number of total users $K$  exceeds the number of  antennas at the BS.
The essence of multiuser diversity was first introduced in \cite{ref:R. Knopp}, where only the strongest user was served by the single antenna BS in each time slot. This work was followed by numerous user selection/scheduling schemes for various scenarios. In \cite{ref:G. Dimic}, a downlink model with multiple single-antenna users and a multiple-antenna BS was considered. The authors analyzed the throughput of the greedy zero-forcing dirty-paper (ZF-DP) algorithm and characterized the probability density function (PDF) of the ordered signal-to-noise ratios (SNRs) of the users. In \cite{ref:A. Goldsmith}, the authors considered zero-forcing beamforming (ZFBF) and proposed an algorithm to schedule the strongest and almost orthogonal users for a similar model. It has been shown that the ZFBF can achieve the same asymptotic sum capacity as that of dirty paper coding. Similar studies on the uplink transmissions were done in \cite{ref:X. Qin}-\cite{ref:J. Choi}. In order to analyze multiuser diversity, extreme value theory has been widely used to provide asymptotic results of throughput \cite{ref:A. Goldsmith}-\cite{ref:M. Sharif Partial}, \cite{ref:M. A. Maddah-Ali}.

The multiple users/antennas selection diversity also plays an important role in improving the \emph{security}. In \cite{ref:M. Pei CL}, the authors investigated the secrecy performance of multiuser scheduling on a  MISO downlink wiretap channel in the presence of a passive multiple-antenna eavesdropper. In \cite{ref:X. Liu}, the authors developed a joint semidefinite programming (SDP) and successive convex approximation (SCA) algorithm to find a feasible subset where legitimate users can satisfy secrecy-outage requirements. Unlike the system models investigated in  \cite{ref:M. Pei CL} and \cite{ref:X. Liu}, the authors in \cite{T. M. Hoang} considered the use of cooperative beamforming and user selection for relay network security. In \cite{ref:J. Lee}-\cite{ref:C Wang SPL}, the authors exploited the multiuser diversity to increase the secrecy degrees of freedom or to improve secrecy rate via jammer selection.  As for MIMO systems, transmit antenna selection can also provide secrecy gains \cite{ref:H. Alves}-\cite{ref:N. Yang}. For a secure transmission system assisted by multiple cooperative relay nodes, the issue of relay selection was addressed in \cite{ref:Z. Ding}, \cite{ref:Selection}-\cite{ref:A. Mabrouk}. However, none of the existing works has studied the relationship between the \emph{ergodic secrecy sum-rate} (ESSR) and the number of the users scheduled concurrently for multiuser uplink transmissions, and neither the scaling law of secure uplink transmissions nor  the exact expression of the multiuser secrecy gain has been presented yet.

\subsection{Main contributions}
In this paper, we investigate secure transmission in a multiuser uplink system, where there are a BS with $M$ receive antennas, $K$ mobile users each with a single antenna, and an eavesdropper with $N$ receive antennas. Our goal is to characterize the achievable ESSR of the system and to reveal its scaling behavior and multiuser secrecy gain via user selection.
Compared with the above related works, our key contributions are summarized as follows:
\begin{enumerate}
\item[1)]	Closed-form expressions of ESSR for a \emph{random} $k$ user selection scheme in both the high and low SNR regimes are presented. We show that the scaling behavior of ESSR with respect to (w.r.t.) the number of served users $k$ is quite different under different system configurations determined by the numbers of the BS antennas and the eavesdropper antennas. When the eavesdropper has the same number of antennas as the BS, the maximum ESSR is achieved at $k=M$ and it scales with $\sqrt{\log M}$  in the high SNR regime. In contrast, the ESSR always grows with $\sqrt{kM}$ in the low SNR regime.

\item[2)]	Low-complexity user selection schemes are proposed for two scenarios i) only  channel state information (CSI) of the main channel (from users to the BS) is available at the BS, referred to as \emph{main CSI}, and ii) both CSIs of the main and wiretap channels are available at the BS, referred to as \emph{full CSI}. Furthermore, closed-form expressions of the achievable ESSRs and  multiuser secrecy gains of these two schemes are also presented in both low and high SNR regimes. We show that multiuser secrecy gain provided by user selection can improve the secrecy performance significantly.

\item[3)]  The impact of the number of antennas at eavesdropper on the ESSR is explored\footnote{In existing works, a widely used assumption is that the number of the eavesdropper antenna is less than that of the BS. When the eavesdropper has more antennas than the BS, it is a more challenging case for secure transmission, which has been seldom studied. }. We show that, when the eavesdropper has a more capable receiver than the BS, serving one user with the strongest channel (TDMA-like) is a favourable secrecy scheme with a multiuser secrecy gain scaling like $\sqrt{{2\log K}}$. Furthermore, we demonstrate that the ESSRs and multiuser secrecy gain are deteriorated by channel estimation errors, especially in the high SNR regime.
   % In the high SNR regime, the existence condition of a positive ESSR for the TDMA-like scheme is $\log \eta \le \sqrt{\frac{2\log K}{M}}$ for main CSI, while it is $\sqrt{\frac{\eta}{1+\eta}}\log \eta \le \sqrt{\frac{2\log K}{M}}$ for full CSI, where $\eta\triangleq \frac{N}{M}$. Similar results of the existence condition are also given for the low SNR regime.
 \end{enumerate}
% In the high SNR regime, it is shown that the multiuser secrecy gain scale with $\sqrt{2\log K}$, for both \emph{main CSI} and \emph{full CSI}.

\subsection{Organization and notations}
The remaining of this paper is organized as follows. In Section II, we describe the system model. Section III gives some mathematical preliminaries and Section VI provides closed-form expressions of the ESSR for the proposed random user selection scheme. Section V proposes two greedy user selection schemes and derives the achievable ESSR of the schemes. Section VI investigates the effect of channel estimation errors on the secrecy performance. Finally, we demonstrate and discuss numerical results in Section VII and conclude our work in Section VIII.

\emph{Notations:} Upper case and lower case bold symbols denote matrices and vectors, respectively. Superscript
${(.)}^\dag$ denotes conjugate transposition, and the notations ${|\mathbf{A}|}$ and
tr(${\mathbf{A}}$) denote the determinant and trace of matrix ${\mathbf{A}}$, respectively. The expectation operator and variance operator are denoted by $\mathbb{E}{(.)}$ and $\mathbb{V}{(.)}$, respectively. The special functions used in this paper are defined as follows, $\{x\}^+ = \max(x,0)$, $\operatorname{erf}(x)=\frac{2}{\sqrt{\pi}}\int_0^{x}e^{-t^2}dt$ is the error function, $\operatorname{Q}(x)=\frac{1}{\sqrt{2\pi}}\int_x^{\infty}e^{-\frac{t^2}{2}}dt$,  $\psi(r)=\sum_{j=1}^{r-1}\frac{1}{j}-\gamma$ is the \emph{digamma} function, and $\gamma= 0.5772\cdots$ is the Euler's constant. The base-$e$ logarithm is denoted by $\log(.)$.

%A technique called user scheduling can be used to improve the system performance of multiuser systems while preserving the advantages of MIMO wireless systems.[A low complexity user scheduling Algorithm for uplink Multiuser MIMO systems]

\section{System Model}
In this work, we consider a wiretap uplink transmission in which a total of $K$ users, each equipped with a single antenna, transmit secrecy signals to a BS equipped with $M$ antennas, wiretapped by an eavesdropper with $N$ antennas. We assume that both the BS and the eavesdropper are equipped with a (not so) large number of antennas\footnote{Under this assumption, we can apply the Gaussian approximation to the mutual information, and present a closed-form expression for the ESSR, which has not been obtained yet. It has been verified in \cite{ref:Z. Wang} and \cite{ref:E. Biglieri} that even for a small number of antennas, the mutual information can be well approximated by a Gaussian distribution.}, i.e., $M,N \ge 10$. In addition, we consider a homogenous scenario where all users to the BS and the eavesdropper have the same average SNR. This scenario is widely adopted for multi-user networks, e.g., in \cite{ref:A. Goldsmith}, \cite{ref:G. Dimic} and \cite{ref:M. A. Maddah-Ali}.
%The channels between each user and the base station as well as the eavesdropper are modeled as zero-mean unit-variance circularly symmetric Gaussian matrixes.We also assume perfect CSI to the legitimate users is available at the BS. This is can be accomplished by sending a pilot signal from each of the users.
Suppose that the served users all transmit independent Gaussian signals with equal average power, when $k$ users are scheduled to transmit simultaneously, the received signals at the BS and at the eavesdropper can be respectively expressed as,
\begin{align}
&\mathbf{y}_b = \sqrt{P}\mathbf{H}\mathbf{s}+\mathbf{n}_b,\\
&\mathbf{y}_e = \sqrt{P}\mathbf{G}\mathbf{s}+\mathbf{n}_e,
\end{align}
where $\mathbf{H} \in \mathbb{C}^{M\times k}$ represents the channel matrix between the BS and  $k$ served users, $\mathbf{G} \in \mathbb{C}^{N\times k} $ represents the channel matrix between the eavesdropper and the $k$ served users, $\sqrt{P}\mathbf{s}$ is a $k \times 1$ vector of symbols simultaneously transmitted by the users (the average transmit power of each user is $P$), $\mathbf{n}_b$ and $\mathbf{n}_e$ are complex Gaussian noise vectors with covariance matrices $\delta_b^2\mathbf{I}$ and $\delta_e^2\mathbf{I}$, respectively. Uncorrelated Rayleigh flat fading channels are considered, and hence the entries of $\mathbf{H}$ and $\mathbf{G}$ are independent and identically distributed (i.i.d.) complex Gaussian random variables with zero mean and unit variance. Without loss of generality, we assume that it holds $\delta_b^2=\delta_e^2=\delta^2$. For notation convenience, we let $\rho\triangleq\frac{P}{\delta^2}$.

Given the above configurations, the achievable sum-rate at the BS and the eavesdropper can be formulated respectively as \cite[pp. 557]{ref:D. Tse}, \cite{ref:Duality Tse}
\begin{align}
&C_b^k(\mathbf{H})=\log\left|\mathbf{I}+\rho\mathbf{H}^{\dagger} \mathbf{H}  \right|, \label{Rate_BS_Per}\\
&C_e^k(\mathbf{G})=\log\left|\mathbf{I}+\rho\mathbf{G}^{\dagger} \mathbf{G}  \right|.
\end{align}

Thus when the BS serves $k$ users simultaneously in each time slot, an achievable secrecy sum-rate is given by  \cite {ref:G. Bagherikaram}
\begin{align}
C_s^{k}&= \left\{C_b^k(\mathbf{H})-C_e^k(\mathbf{G})\right\}^+ \nonumber \\
&=\left\{\log\left|\mathbf{I}+\rho\mathbf{H}^{\dagger} \mathbf{H}  \right|-\log\left|\mathbf{I}+\rho\mathbf{G}^{\dagger} \mathbf{G}  \right|\right\}^+.  \label{ISR_de}
\end{align}

The goal of this paper is to find a scaling law to demonstrate the relationship between the secrecy sum-rate and the number of the served users $k$ as well as the number of the total users $K$. To quantify the scaling law of the scenario described above, we assume that both the main channels and wiretap channels are ergodic block-fading \cite{ref:M. Pei}. We further assume a scenario with delay-tolerant traffic, and use ESSR as the performance metric, which is defined as \cite{ref:Gopala}, \cite{ref:Y. Liang}
\begin{align}
R_s^{k}= \mathbb{E}\left[\left\{\log\left|\mathbf{I}+\rho\mathbf{H}^{\dagger} \mathbf{H}  \right|-\log\left|\mathbf{I}+\rho\mathbf{G}^{\dagger} \mathbf{G}  \right|\right\}^+\right].  \label{ESSR_de1}
\end{align}
%We will give a closed-form expression of the ESSR in Section-IV. Then the multiuser secrecy gain is characterized in Section-V.

\textbf{\emph{Remark 1}}: A lower bound of \eqref{ESSR_de1} can be given as
\begin{align}
{R}_{s,low}^{k}= \left\{\mathbb{E}\left[\log\left|\mathbf{I}+\rho\mathbf{H}^{\dagger} \mathbf{H}  \right|-\log\left|\mathbf{I}+\rho\mathbf{G}^{\dagger} \mathbf{G}  \right|\right]\right\}^+,  \label{ESSR_de2}
\end{align}
by applying Jensen's inequality.
In most cases, \eqref{ESSR_de2} is mathematically more convenient for analyzing than \eqref{ESSR_de1} since the non-linear operator $\{\cdot\}^+$ is taken after the expectation, and it is widely adopted alternatively as the performance metric, e.g., in \cite{ref:T. Shang-Ho}-\cite{ref:X. Zhou Achievablerate}, and \cite{ref:M. Pei}. However, when the eavesdropper has more antennas than the BS, it holds that ${R}_{s}^{k} > 0$ and ${R}_{s,low}^{k} = 0$. Hence, the expression in \eqref{ESSR_de2} can not characterize the actual ESSR in such a case.  So far, a closed-form expression of \eqref{ESSR_de1} has not been obtained yet.

\section{Mathematical Preliminaries}
Before proceeding, we provide three lemmas that will be used in the following analytical derivations.
\begin{lemma}
Let $C(\mathbf{H})\triangleq \log\left|\mathbf{I}+\rho\mathbf{H}^{\dagger} \mathbf{H}  \right|$, where $\mathbf{H} \in \mathbb{C}^{M\times k}$. In the high SNR regime ($\rho \rightarrow \infty$), the distribution of the channel capacity $C(\mathbf{H})$ is approximately given as
\begin{align}
C(\mathbf{H}) \sim \mathcal{N}\left(\mu_{Mk},\sigma_{Mk}^2\right),  \label{Lemma_1}
\end{align}
where $\mu_{M,k}$ and $\sigma_{M,k}^2$ are given as follows
\begin{align}
\mu_{M,k}=\left\{
\begin{aligned}
&k\log\rho+ \sum_{i=1}^{k} \psi(M-i+1),\ \ \ \ k \le M, \\
&M\log\rho+\sum_{i=1}^{M}\psi(k-i+1), \ \ \ \ k > M, \label{Mu_Mk}
\end{aligned}
\right.
\end{align}
\begin{align}
\sigma_{M,k}^2=\left\{
\begin{aligned}
&\sum_{i=1}^{k-1}\frac{i}{(M-k+i)^2}+\frac{k}{M},\ \ \ \ \ \ \ k \le M, \\
&\sum_{i=1}^{M-1}\frac{i}{(k-M+i)^2}+\frac{M}{k}, \ \ \ \ \ \ k > M. \label{Sigma_Mk}
\end{aligned}
\right.
\end{align}

\end{lemma}

\begin{IEEEproof}
When $k \le M$ and $M$ is sufficiently large, using Theorem 3 given in \cite{ref:B. M. Hochwald}, we readily have that $C(\mathbf{H})$ is approximately a Gaussian random variable with mean and variance given by, respectively,
\begin{align}
\mu_{M,k}&=k\log\rho+ \sum_{i=1}^{k} \psi(M-i+1),  \\
\sigma_{M,k}^2&=\sum_{i=1}^{k-1}\frac{i}{(M-k+i)^2}+k\left(\frac{\pi^2}{6}-\sum_{i=1}^{M-1}\frac{1}{i^2}\right). \label{Or_sigma}
\end{align}

Resorting to \cite[Eqn. 9.521]{ref:Table}, the following approximation holds when $M$ goes large,
\begin{align}
\frac{\pi^2}{6}-\sum_{i=1}^{M-1}\frac{1}{i^2} \approx \frac{1}{M}.  \label{1/M}
\end{align}

Substituting \eqref{1/M} into \eqref{Or_sigma} yields the first expression in \eqref{Sigma_Mk}. Similarly, we can obtain the result for the case $k > M$.
\end{IEEEproof}
It is worth to point out that $C(\mathbf{H})$ is very well approximated by a Gaussian random variable
even for a small number of antennas $M$ \cite{ref:Z. Wang}, \cite{ref:E. Biglieri}.
%For any set of independent identical distribution (i.i.d) random variables $X_1,X_2,\cdots,X_K$ , denoting $X_{(m)}= \max(X_1,X_2,\cdots,X_K)$.
\begin{lemma}
When $Y^K_{(1)}$ is the maximum of a sequence of $K$ i.i.d. random variables following $\mathcal{N}(\mu,\sigma^2)$ distribution, as $K \rightarrow \infty$, it satisfies
\begin{align}
\underset{K\rightarrow \infty}{\lim}\mathbb{P}\left(Y^K_{(1)}\le a_K t+b_K \right)=e^{-e^{-t}},
\end{align}
where
\begin{align}
&a_K =\frac{\sigma}{\sqrt{2\log K}}, \label{MAX_con_a}  \\
&b_K =\sigma\sqrt{2\log K} -\sigma \frac{\log(4\pi \log K)}{2\sqrt{2 \log K}}+\mu.  \label{MAX_con_b}
\end{align}

The mean of $Y^K_{(1)}$ approaches
\begin{align}
\mathbb{E}\left[Y^K_{(1)}\right]&=\sigma\sqrt{2\log K}  -\sigma \frac{\log(4\pi \log K)-2\gamma}{2\sqrt{2 \log K}}+\mu.
\end{align}
\end{lemma}
\begin{IEEEproof}
Please see Appendix A.
\end{IEEEproof}

\begin{lemma}
When $Y_{(r)}^K$ is the $r$-th largest of a sequence of $K$ i.i.d. random variables following $\mathcal{N}(\mu,\sigma^2)$  distribution, as $K \rightarrow \infty$, its mean would approach
\begin{align}
\mathbb{E}[Y_{(r)}^K]&=\sigma\sqrt{2\log K}  -\sigma \frac{\log(4\pi \log K)+2\psi(r)}{2\sqrt{2 \log K}}+\mu.
\end{align}
\end{lemma}
\begin{IEEEproof}
Please see Appendix B.
\end{IEEEproof}

\section{ESSR Scaling Law  under \\ Random User Selection}
Let us now study the ESSR scaling law of \emph{random} user selection.
In a random user selection scheme, the BS randomly selects $k$ users to serve in each time slot. In a $k$-user uplink transmission without secrecy constraints, it is shown that the sum capacity increases monotonically with $k$ \cite[pp. 565]{ref:D. Tse}. Especially, in the high SNR regime, the sum capacity only grows logarithmically w.r.t. $k$ as $k$ increases beyond $M$ \cite[pp. 565]{ref:D. Tse}. However,  the secrecy sum-rate is the rate difference between the BS and the eavesdropper, thus the aforementioned results can not be generalized directly to secrecy communications. Since the system performance behaves differently in the high and low SNR regimes, herein we consider the two special regimes respectively in the following subsections.

\subsection{High SNR Regime}
In the high SNR regime ($\rho \rightarrow \infty$), we provide a closed-form expression of ESSR in the following theorem.
\begin{theorem}
For a given $k$, the achievable ESSR of random $k$ user selection in the high SNR regime is approximated as
\begin{align}
R_s^{\operatorname{k\_ran}} \approx \frac{\sigma_k}{\sqrt{2\pi}}e^{-\frac{\mu_k^2}{2\sigma_k^2}}+\frac{\mu_k}{2}\left(1+\operatorname{erf}\left(\frac{\mu_k}{\sqrt{2} \sigma_k}\right)\right), \label{Theorem1}
\end{align}
where
\begin{align}
\mu_k&=\mu_{M,k}-\mu_{N,k}, \label{mu_k}\\
\sigma_k^2&= \sigma_{M,k}^2 + \sigma_{N,k}^2,  \label{sigma_k}
\end{align}
with $\mu_{M,k}$, $\mu_{N,k}$, $\sigma_{M,k}$, and $\sigma_{N,k}$ given in Lemma 1.
\end{theorem}

\begin{IEEEproof}
From the results given in Lemma 1, we have $C_b^k(\mathbf{H})\sim \mathcal{N}\left(\mu_{M,k},\sigma_{M,k}^2\right)$ and $C_e^k(\mathbf{G})\sim \mathcal{N}\left(\mu_{N,k},\sigma_{N,k}^2\right)$ in the high SNR regime. Let $Z = C_b^k(\mathbf{H})-C_e^k(\mathbf{G})$, which obviously obeys $Z\sim \mathcal{N}\left(\mu_k,\sigma_k^2\right)$. Recalling to \eqref{ESSR_de1}, the ESSR of random $k$ user selection can be calculated as
\begin{align}
R_s^{\operatorname{k\_ran}}&=\mathbb{E}\left\{Z\right\}^+ \nonumber \\
&\approx\int_0^{\infty}\frac{z}{\sqrt{2\pi \sigma_k^2}}e^{-\frac{(z-\mu_k)^2}{2\sigma_k^2}}dz \nonumber \\
&=\frac{\sigma_k}{\sqrt{2\pi}}e^{-\frac{\mu_k^2}{2\sigma_k^2}}+\frac{\mu_k}{2}\left(1+\operatorname{erf}\left(\frac{\mu_k}{\sqrt{2} \sigma_k}\right)\right). \nonumber
\end{align}

The proof is completed.
\end{IEEEproof}

\textbf{\emph{Remark 2}}: Although Theorem 1 considers the $k$-user uplink scenario, it can be applied directly to a point-to-point  MIMO wiretap channel with $k$ transmit antennas, $M$ receive antennas and an eavesdropper with $N$ antennas.

Using Theorem 1, the scaling law of ESSR w.r.t. the number of the served users $k$ can be investigated in more details. Note that $M<N$ implies that the eavesdropper has a more capable receiver than the BS, and vice versa. We will show that the scaling laws are different for the cases with a more capable eavesdropper or a less capable eavesdropper. Therefore, the relationship between the ESSR and $k$ for the cases $M > N $, $M=N$ and $M < N$ is explored in the following three corollaries, respectively.

%%We can obtain the optimal served number of the served users in the sense with the maximum ESSR.
%%In our work, we consider the K-user uplink and focus on the ergodic sum secrecy rate.

\textbf{Corollary 1.} For the case $M >N$, $R_s^{\operatorname{k\_ran}}$ does not rely on $\rho$ as $k \le N$. While for $k > N$, it holds that $R_s^{\operatorname{k\_ran}} \thickapprox \mu_k$.

\begin{IEEEproof}
When $M >N$, $\mu_k$ in \eqref{mu_k} can be rewritten as
\begin{align}
\mu_k = \begin{aligned}
\begin{cases}
\sum_{i=1}^{k}\Big[\psi(M-i+1)-\psi(N-i+1)\Big],&k \le N,  \\
\sum_{i=1}^{k}\psi(M-i+1)-\sum_{i=1}^{N}\psi(k-i+1) \\
 \hspace{2cm} + (k-N)\log \rho,&\hspace{-0.75cm} N < k \le M,\\
\sum_{i=N+1}^{M}\psi(k-i+1)+(M-N)\log \rho,& k > M.   \\
\end{cases}
\end{aligned}\nonumber   %这个点一定不能少。。。,与前面的括号对应
\end{align}

We see that when $k \le N$, neither $\mu_k$ nor $\sigma_k^2$ is a function of $\rho$, hence $R_s^{\operatorname{k\_ran}}$ in \eqref{Theorem1} does not rely on $\rho$.
%we have $\sigma_k^2= \sum_{i=1}^{k-1}\frac{i}{(M-k+i)^2}+\sum_{i=1}^{N-1}\frac{i}{(k-N+i)^2}+\frac{k}{M}+\frac{N}{k}$.

When $k > N$, it holds that $\frac{\mu_k}{\sqrt{2} \sigma_k} \rightarrow \infty$ as $\rho \rightarrow \infty$. For a sufficient large $x$, error function $\operatorname{erf}(x)$ can be approximated by $\operatorname{erf}(x) \thickapprox 1- \frac{e^{-x^2}}{\sqrt{\pi}x}$. Therefore, we have
\begin{align}
R_s^{\operatorname{k\_ran}} &\approx \frac{\sigma_k}{\sqrt{2\pi}}e^{-\frac{\mu_k^2}{2\sigma_k^2}}+\frac{\mu_k}{2}\left(1+\operatorname{erf}\left(\frac{\mu_k}{\sqrt{2} \sigma_k}\right)\right) \nonumber \\
&\thickapprox \frac{\sigma_k}{\sqrt{2\pi}}e^{-\frac{\mu_k^2}{2\sigma_k^2}} + \frac{\mu_k}{2}\left(2-\frac{\sqrt{2}\sigma_k}{\sqrt{\pi}\mu_k}e^{-\frac{\mu_k^2}{2\sigma_k^2}}\right) \nonumber \\
&=\mu_k.
\end{align}

The proof is completed.
\end{IEEEproof}

Corollary 1 tells that when $M>N>k$, increasing the transmit power will not improve the ESSR. When $k>N$, the ESSR is solely determined by $\mu_k$, which leads to two interesting observations:
\begin{enumerate}
\item[1)]	When $N < k \le M$, the first two terms in $\mu_k$ grow sub-linearly with $k$, thus $\mu_k$ can be rewritten as $\mu_k = k\log \rho + o(k)-N\log \rho$, which implies the ESSR grows linearly with $\log\rho$ as $k$ increases.
\item[2)]	However, as $k > M$, it holds that $\mu_k \le (M-N-1)\log(k-N-1)+(M-N)\log \rho$ \cite{ref:R. M. Young}, thus $\mu_k$ will not grow faster than $\log(k-N-1)$ w.r.t. $k$. Therefore, increasing $k$ does not improve the ESSR as fast as that of the case $N < k \le M$.
 \end{enumerate}

We can see that the scaling behavior of the ESSR differs significantly from the case without secrecy constraints. These two observations will be illustrated clearly later in Fig. 1.

\textbf{Corollary 2.} For the case $M = N$, the maximum ESSR is achieved at $k=M$, which can be well approximated by
\begin{align}
R_s^{\operatorname{M\_ran}} \approx \sqrt{\frac{1}{\pi}\Big(\log (M-1) +\gamma + 1\Big)}. \label{Coro_2}
 \end{align}

\begin{IEEEproof}
Please see Appendix C.
\end{IEEEproof}

Corollary 2 indicates that,  if the eavesdropper has an equal number of antennas as the BS, the maximum ESSR is achieved when the BS serves $M$ users simultaneously. It is shown that $R_s^{\operatorname{M\_ran}}$ scales with $\sqrt{\log M}$, while the sum-capacity grows linearly with $M$ in conventional systems without secrecy constraints.

\textbf{Corollary 3.} For the case $M < N$, it holds that
 \begin{align}
R_s^{\operatorname{k\_ran}} \le \frac{\sigma_k}{\sqrt{2\pi}}e^{-\frac{\mu_k^2}{2\sigma_k^2}}.  \label{Coro3}
\end{align}
\begin{IEEEproof}
From \eqref{mu_k}, we know that $\mu_k < 0$ as $M<N$. Meanwhile, when $x<0$, it holds that $1+\operatorname{erf}(x) = 1- \operatorname{erf}(|x|) > 0 $. Therefore, the second term in \eqref{Theorem1} is always negative, then we can obtain the result in \eqref{Coro3}.
\end{IEEEproof}

\begin{figure}[t]
\begin{center}
\includegraphics[width=2.8in]{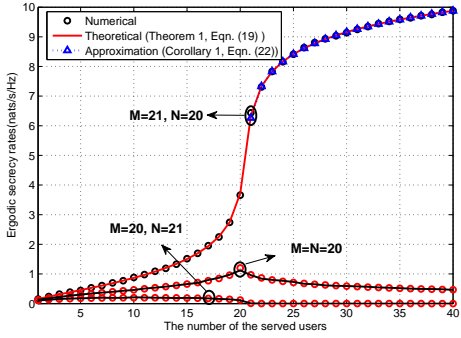}
\end{center}
\vspace{-0.3cm}
\caption{Comparisons among the theoretical, simulation, and approximation results in the high SNR regime.}
\end{figure}

Corollary 3 provides an upper bound for the ESSR. When $M<k<N$, we have $\mu_{M,k+1}-\mu_{M,k} = \psi(k+1) - \psi(k+1-M)$ and $\mu_{N,k+1}-\mu_{N,k} = \log \rho + \psi(N-k)$. As can be seen, increasing $k$ does not significantly improve the sum-rate for the BS. This is because serving one more user would not only increase the sum-rate at the BS but also bring interference to other users. In contrast, the sum-rate achieved at the eavesdropper has a great increment as $k$ increases. Since the secrecy rate is the rate difference between the rate of the main channel and that of the wiretap channel, $R_s^{\operatorname{k\_ran}}$ would decrease as $k>M$. Moreover, $\mu_k$ scales with $-\log \rho$ as $k>M$, $\mu_k^2$ becomes sufficiently large and hence $R_s^{\operatorname{k\_ran}}$ would quickly converge to zero. Besides,  we can find that the upper bound would converge to zero as $N$ is much greater than $M$, and thus a positive ESSR can not be achieved.

Theorem 1 and the Corollaries 1-3 demonstrate the scaling behaviors of the ESSR w.r.t. the number of random selected user $k$ in the high SNR regime.  We can see that the scaling behavior depends heavily on the relationship between the number of antennas at the BS and the eavesdropper.

To verify these analytical results, we perform Monte Carlo experiments each with 10000 independent trials to obtain the numerical results of \eqref{ESSR_de1} at a desired SNR $\rho = 30\ dB$. The results are plotted in Fig. 1. It is shown that the theoretical results are consistent with the numerical results even with moderate numbers of antennas at the eavesdropper and the BS. When the BS has one more antenna than the eavesdropper ($M=N+1$), $R_s^{\operatorname{k\_ran}}$ is well approximated by $\mu_k$ as $k>N$ and a significant improvement of ESSR happens as $k=N+1$. In contrast, adding one more antenna to the eavesdropper ($M=N-1$), the ESSR becomes much lower than those of the cases $M = N$ and $M=N+1$. As Corollary 3 has stated, the ESSR quickly converges to zero when $k > M$. The figure also validates that the maximum ESSR is achieved at $k = M$ when the eavesdropper has an equal number of antennas as the BS ($M=N$).

\subsection{Low SNR Regime}
In the low SNR regime ($\rho \rightarrow 0$), we have
\begin{align}
C_b^{k}&= \log\left|\mathbf{I}+\rho\mathbf{H}^{\dagger} \mathbf{H}  \right|  =\sum_{i=1}^{k}\log(1+\rho\lambda_{bi}) \nonumber \\
& = \rho \sum_{i=1}^{k}\lambda_{bi} + {o}(\rho^2)  \approx \rho\operatorname{tr}(\mathbf{H}^{\dagger} \mathbf{H}), \label{rho-0}
\end{align}
where $\lambda_{bi}$, $i=1,\cdots,k$, are the $k$ non-zero eigenvalues of $(\mathbf{H}^{\dagger} \mathbf{H})$. Since the entries of $\mathbf{H}$ are complex Gaussian variables, it holds that $\operatorname{tr}(\mathbf{H}^{\dagger} \mathbf{H}) \sim \chi_{2Mk}^2$ \cite[pp. 742]{J. Proakis}. Then after some manipulations, we can obtain a closed-form approximation of the ESSR in the low SNR regime. However, it can not offer insights into the relationship between the ESSR and the number of the served users.  Therefore, we use the Gaussian approximation of the Chi-square to provide a more insightful expression.

Let $Z= \mathbf{h_i}^\dag\mathbf{h_i}$, where $\mathbf{h_i}$ is the $i$-th column of $\mathbf{H}$. It holds $Z \sim \chi_{2M}^2$ \cite[pp. 742]{J. Proakis}, and both the mean and variance of $Z$ are $M$. Using the central limit theorem, the CDF of the variable $Z$ can be approximated as
\begin{align}
\underset{M\rightarrow \infty}{\lim}\mathbb{P}\left(Z \ge t\right)= \operatorname{Q}\left(\frac{t-M}{\sqrt{M}}\right). \label{Q(x)_App}
\end{align}

It has been proved in \cite{ref:D. Horgan} that the error resulting from the approximation of \eqref{Q(x)_App} is bounded by $\frac{1}{\sqrt{9\pi M/2}}$. Supposing that $M=20$, the error is about $0.06$, which is fairly small. Therefore the PDF of  $Z$ can be approximated by a normal distribution $\mathcal{N}{(M, M)}$ even when $M$ is not very large. We present the following theorem to approximate the ESSR in the low SNR regime.

\begin{theorem}
For a given $k$, the ESSR of random $k$ user selection in the low SNR regime can be approximated as
\begin{align}
R_s^{\operatorname{k\_ran}} \approx \frac{\beta_k\rho}{\sqrt{2\pi}}e^{-\frac{\alpha_k^2}{2\beta_k^2}}+\frac{\alpha_k\rho}{2}\left(1+\operatorname{erf}\left(\frac{\alpha_k}{\sqrt{2} \beta_k}\right)\right), \label{Rs_Low_App}
\end{align}
where $\alpha_k = k(M-N)$ and $\beta_k^2 = k(M+N)$.
\end{theorem}

\begin{IEEEproof}
In the low SNR regime, as $\rho \rightarrow 0$, we have
$C_s^{k}\thickapprox\rho\left\{\operatorname{tr}(\mathbf{H}^{\dagger} \mathbf{H})-\operatorname{tr}(\mathbf{G}^{\dagger} \mathbf{G})\right\}^+$. Let $X=\operatorname{tr}(\mathbf{H}^{\dagger} \mathbf{H})$ and $Y=\operatorname{tr}(\mathbf{G}^{\dagger} \mathbf{G})$. It holds that $X \sim \chi_{2Mk}^2$  and $Y \sim \chi_{2Nk}^2$. As we have discussed previously, a Chi-square random variable can be approximated by a Gaussian one. Therefore, $X-Y$ is distributed according to $\mathcal{N}{(\alpha_k, \beta_k^2)}$.
Similar to the proof of Theorem 1, we can readily obtain the result in \eqref{Rs_Low_App}.
\end{IEEEproof}
Theorem 2 shows that the ESSR always increases with $\rho$ in the low SNR regime. The relationship between the ESSR and the number of the BS antennas as well as the number of the eavesdropper antennas for the cases $M>N$ and $M<N$ can be analyzed similarly as that in the high SNR regime.  An interesting and insightful result for the special case $M=N$ is presented in the following Corollary.

\textbf{Corollary 4.} For a given $k$, as $M=N$, the ESSR of random $k$ user selection in the low SNR regime is approximated as
\begin{align}
R_s^{\operatorname{k\_ran}} \approx\rho\sqrt \frac{kM}{\pi}.
\end{align}

Different from Corollary 2, the ESSR always grows with the number of served users. As $k = M$, the ESSR scales with $M$ in the low SNR regime while scales with $\sqrt{\log{M}}$ in the high SNR regime. With this observation, we know that the increase of the BS antennas can offer more secrecy benefits in the low SNR regime. Besides, we should note that the ESSR monotonically increases with $\rho$ in the low SNR regime as the conventional systems without secrecy constraints.

\begin{figure}[t]
\begin{center}
\includegraphics[width=2.8in]{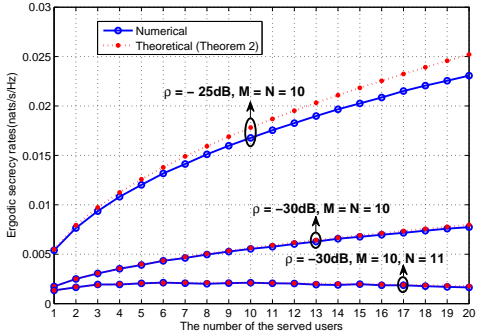}
\end{center}
\vspace{-0.3cm}
\caption{Comparisons between the theoretical and numerical results in the low SNR regime.}
\end{figure}

Fig. 2 depicts the results given in Theorem 2 and Corollary 4.  The numerical results of \eqref{ESSR_de1} are also obtained by performing Monte Carlo experiments. We can see that the theoretical results agree well with the behavior of numerical results when $\rho=-30\ dB$. Fig. 2 also verifies that the ESSR grows with $\rho$ in the low SNR regime.

%Although the analytical results are not very accurate as $\rho$ gets large, it can be seen that the scaling behavior with the number of served users is still consistent.

\subsection{Large Scale Analysis for Random User Selection}
MIMO systems with very large antenna arrays at the BS, so called massive MIMO systems, is one of the key technologies to improve spectral-energy efficiency for future wireless communications. Recently, there has been a great deal of interest in multiuser massive MIMO systems \cite{ref:H. Q. Ngo}. In order to demonstrate the potential of massive MIMO to enhance security, we give a large scale analysis of the ESSR for random user selection.
\begin{theorem}
In a large scale system, for a fixed $k$, the ESSR of the random user selection is approximately given by
\begin{align}
R_{s,lar}^{\operatorname{k\_ran}}\approx \left\{k\log\frac{1+M\rho}{1+N\rho}\right\}^+.
\end{align}
\end{theorem}

\begin{IEEEproof}
Using the results of \cite[Theorem 1]{ref:B. M. Hochwald} and Theorem 2, we can easily complete the proof.
\end{IEEEproof}

In a large scale multiple antenna system, as the number of antennas grows, the channel quickly ``hardens'', in
the sense that the mutual information converges to its mean. The phenomenon of channel hardening makes communications secure only when the BS has more antennas than the eavesdropper, which can be seen from Theorem 3.

\section{Greedy User Selection and \\ Multiuser Secrecy Gain}
In Section IV, we consider a random user selection in secure communications and show that the ESSR would decrease significantly even when the eavesdropper has only one more antenna than the BS. Obviously, the eavesdropper has much more antennas than the BS is a challenging scenario for a secure transmission. Fortunately, in an uplink transmission with a large number of users, the BS can enhance security by selecting the best set of users to communicate with in each time slot, resulting in the multiuser secrecy gain.  In order to highlight the secrecy improvement achieved by user selection, we focus on the case $N \ge M$ in this section, i.e., the eavesdropper is a more capable receiver than the BS. As shown in Section III, as $k > M$, increasing $k$ results in decrease of ESSR when $N \ge M$. Hence, the case $k \le M$ is of our interest.

Note that the user selection scheme is different for various CSI assumptions, i.e., \emph{main CSI} case or \emph{full CSI} case. Typically the instantaneous CSI of the eavesdropper is not available. Here, we consider the full CSI case to investigate the secrecy gain obtained from the prior knowledge of the eavesdropper's CSI, which can be taken as a benchmark or a secrecy performance upper bound to evaluate the multiuser gain achieved by user selection. %Hence the full CSI case is taken into account.
%In practice, the eavesdropper may be a false BS, thus the users can track the eavesdropper's CSI and then feedback them to the BS.

The optimal user selection for \emph{main CSI} and \emph{full CSI} are formulated in the following, respectively.

Let $U=\{1,2,\cdots,K\}$ denote the set of all $K$ users, and let $S_{k} = \{s_1,s_2,\cdots,s_k\}$ denote the set of $k$ selected/served users. For the \emph{main CSI} case, the user selection problem is formulated as follows: Given $\mathbf{H} \in \mathbb{C}^{M \times K}$, select a set of channels $\mathbf{H}(S_k)=[\mathbf{h}_{s_1},\mathbf{h}_{s_2},\cdots,\mathbf{h}_{s_k}]$ such that the sum-rate achieved at the BS is maximized, i.e.,
\begin{align}
C_{s}^{\operatorname{Mk\_opt}}= &\Big\{\Big[\underset{S_{k}}{\max} \ \ {C_b^k\left(\mathbf{H}(S_k)\right)}\Big]-C_e^k(\mathbf{G}(S_k))\Big\}^+ \nonumber \\
=&\Big\{\Big[\underset{S_{k}}{\max} \ \ \log\left|\mathbf{I}_k+\rho\mathbf{H}(S_k)^{\dagger} \mathbf{H}(S_k)  \right|\Big]  \nonumber \\
&-\log\left|\mathbf{I}_k+\rho\mathbf{G}(S_k)^{\dagger} \mathbf{G}(S_k)  \right|\Big\}^+.   \label{Cs_ik}
\end{align}

For the \emph{full CSI} case, the user selection problem is formulated as follows: Given $\mathbf{H} \in \mathbb{C}^{M \times K}$ and $\mathbf{G} \in \mathbb{C}^{N \times K}$, select a set of users with the corresponding channels $\mathbf{H}(S_k)=[\mathbf{h}_{s_1},\mathbf{h}_{s_2},\cdots,\mathbf{h}_{s_k}]$ and $\mathbf{G}(S_k)=[\mathbf{g}_{s_1},\mathbf{g}_{s_2},\cdots,\mathbf{g}_{s_k}]$ such that the secrecy sum-rate achieved is maximized, i.e.,
\begin{align}
C_{s}^{\operatorname{Fk\_opt}}
&=\Big\{\underset{S_{k}}{\max} \ \ \Big[{C_b\left(\mathbf{H}(S_k)\right)}-C_e(\mathbf{G}(S_k)) \Big] \Big\}^+ \nonumber \\
&=\left\{\underset{S_{k}}{\max} \ \ \log \frac {\left|\mathbf{I}_k+\rho\mathbf{H}(S_k)^{\dagger} \mathbf{H}(S_k)  \right|}  {\left|\mathbf{I}_k+\rho\mathbf{G}(S_k)^{\dagger} \mathbf{G}(S_k)  \right|}\right\}^+.  \label{Cs_iik}
\end{align}

The problems given in \eqref{Cs_ik} and \eqref{Cs_iik} can be solved by exhaustive search. For a given $k$, traverse all possible $k-$tuples $S_k$ and select a set to maximize $C_b\left(\mathbf{H}(S_k)\right)$ or $C_{s}^{\operatorname{Fk}}(\mathbf{H}(S_k),\mathbf{G}(S_k))$. However, such an exhausted search has a prohibitive complexity, which is not practical for a large-scale network. In the following, we present two greedy user selection algorithms with low complexity, in the high and low SNR regimes, respectively \footnote{It is to be pointed out that we consider homogeneous channel conditions in this work. Although our proposed algorithms schedule the optimal set of users to maximize secrecy sum-rate at each time slot, it is fair for each user, since all users have the same distribution of SNR. The fairness for heterogeneous case in which different users have different average SNR values will be investigated in future study.}. Furthermore, we provide comprehensive ESSR analysis for both algorithms. To characterize the impact of user selection on secrecy, we define the multiuser secrecy gain as
\begin{align}
\Delta_s^{\operatorname{k\_opt}} \triangleq &\mathbb{E}\Big[C_b^{\operatorname{k\_opt}}(\mathbf{H})-C_e^{\operatorname{k\_opt}}(\mathbf{G})\Big] \nonumber \\
&-\mathbb{E}\Big[C_b^{\operatorname{k\_ran}}(\mathbf{H})-C_e^{\operatorname{k\_ran}}(\mathbf{G})\Big].
\end{align}

As can be seen, the multiuser secrecy gain is defined as the ESSR difference between the optimal user selection and the random user selection. Therefore, it indicates the benefit from multiuser gain for secure communications.

\subsection{High SNR Regime}
In the high SNR regime, the sum-rate achieved at the BS can be calculated as \cite{ref:G. Dimic}
    \begin{align}
&C_b^k\left(\mathbf{H}(S_k)\right) \nonumber \\
=&  \log\left|\mathbf{I}_k+\rho\mathbf{H}(S_k)^{\dagger} \mathbf{H}(S_k)  \right| \nonumber \\
 \overset{(a)}= &  \log\left|\mathbf{H}(S_k)^{\dagger} \mathbf{H}(S_k)  \right| + k\log\rho+{o}(1/\rho)  \nonumber \\
 \approx& \log\left|[\mathbf{H}(S_{k-1})\ \ \mathbf{h}_{s_k}]^{\dagger} [\mathbf{H}(S_{k-1}) \ \ \mathbf{h}_{s_k}]  \right| + k\log\rho  \nonumber \\
=& \log \left| \begin{array}{cc}
\mathbf{H}(S_{k-1})^{\dagger} \mathbf{H}(S_{k-1}) & \mathbf{H}(S_{k-1})^{\dagger}\mathbf{h}_{s_k}\\
\mathbf{h}_{s_k}^{\dagger}\mathbf{H}(S_{k-1}) & \mathbf{h}_{s_k}^{\dagger}\mathbf{h}_{s_k}
\end{array}\right| + k\log\rho\nonumber
\\ \overset{(b)}=& \log \left\{\left|\mathbf{H}(S_{k-1})^{\dagger} \mathbf{H}(S_{k-1})\right|\left|\mathbf{h}_{s_k}^{\dagger}\mathbf{A}_{k-1}^{\bot}\mathbf{h}_{s_k} \right|\right\}+ k\log\rho \nonumber \\
=&  \log \left|\mathbf{H}(S_{k-1})^{\dagger} \mathbf{H}(S_{k-1})\right| +\log\left|\mathbf{h}_{s_k}^{\dagger}\mathbf{A}_{k-1}^{\bot}\mathbf{h}_{s_k} \right|+ k\log\rho \nonumber \\
=&  \sum_{l=1}^{k}\log\left|\mathbf{h}_{s_l}^{\dagger}\mathbf{A}_{l-1}^{\bot} \mathbf{h}_{s_l} \right| + k\log\rho, \label{MMSE_SIC_Cb}
\end{align}
where step $(a)$ follows Eqn. (67) in \cite{ref:B. M. Hochwald}, step $(b)$ follows Eqn. (6.2.1) in \cite{ref:Meyer}, $\mathbf{H}(S_{l})=[\mathbf{h}_{s_1},\cdots,\mathbf{h}_{s_{l-1}}]$, and $\mathbf{A}_{l-1}^{\bot} =\mathbf{I}_M - \mathbf{H}(S_{l-1})\left(\mathbf{H}(S_{l-1})^{\dagger} \right.\left.
 \mathbf{H}(S_{l-1})\right)^{-1}\mathbf{H}(S_{l-1})^{\dagger}$. Obviously, it holds that $\mathbf{H}(S_{0}) = \mathbf{0}$. Similar to \eqref{MMSE_SIC_Cb}, the instantaneous secrecy sum-rate for the full CSI case can be approximated as
\begin{align}
C_{s}^{\operatorname{Fk}}&=\left\{\log \frac {\Big|\mathbf{I}+\rho\mathbf{H}(S_k)^{\dagger} \mathbf{H}(S_k)  \Big|}  {\Big|\mathbf{I}+\rho\mathbf{G}(S_k)^{\dagger} \mathbf{G}(S_k)  \Big|} \right\}^+  \nonumber \\
&\approx \left\{\sum_{l=1}^{k}\log \frac{\Big|\mathbf{h}_{s_l}^{\dagger}\mathbf{A}_{l-1}^{\bot} \mathbf{h}_{s_l} \Big|}{{\Big|\mathbf{g}_{s_l}^{\dagger}\mathbf{B}_{l-1}^{\bot} \mathbf{g}_{s_l} \Big|}}\right\}^+, \label{MMSE_SIC_Rs}
\end{align}
where $\mathbf{G}(S_{l})=[\mathbf{g}_{s_1},\cdots,\mathbf{g}_{s_{l}}]$, and $\mathbf{B}_{l-1}^{\bot} =\mathbf{I}_N-\mathbf{G}(S_{l-1})\left(\mathbf{G}(S_{l-1})^{\dagger}\mathbf{G}(S_{l-1})\right)^{-1}\mathbf{G}(S_{l-1})^{\dagger}$.
\begin{algorithm}[t]
\caption{Greedy user selection for the \emph{main CSI} case.}
Step 1) Initialization:
\begin{itemize}
\item[-] Set $l$ = 1.
\item[-] Select a user such that $s_1=\arg \underset{j\in U} {\max}\ \ |\mathbf{h}_j^{\dagger}\mathbf{h}_j|$.
\item[-] Set $S_1$ = {$s_1$}.
\end{itemize}
Step 2) While $l \le k$, select the $l$th user as follows:
\begin{itemize}
\item[-] Increase $l$ by 1.
\item[-] Select a user $s_l$ such that $s_l=\arg \underset{j\in U} {\max} \ \ |\mathbf{h}_j^{\dagger}\mathbf{A}_{l-1}^{\bot} \mathbf{h}_j|$.
\item[-] Set $S_l$ = $S_{l-1} \cup s_l$.
\end{itemize}
\end{algorithm}
\begin{algorithm}[t]
\caption{Greedy user selection for the \emph{full CSI} case.}
Step 1) Initialization:
\begin{itemize}
\item[-] Set $l$ = 1.
\item[-] Select a user such that $s_1=\arg \underset{j\in U} {\max}\ \ \frac{|
\mathbf{h}_j^{\dagger}\mathbf{h}_j|}{|\mathbf{g}_j^{\dagger}\mathbf{g}_j|}$.
\item[-] Set $S_1$ = {$s_1$}.
\end{itemize}
Step 2) While $l \le k$, select the $l$th user as follows:
\begin{itemize}
\item[-] Increase $l$ by 1.
\item[-] Select a user $s_l$ such that $s_l=\arg \underset{j\in U} {\max} \ \ \frac{| \mathbf{h}_j^{\dagger}\mathbf{A}_{l-1}^{\bot}\mathbf{h}_j|}
{|\mathbf{g}_j^{\dagger}\mathbf{B}_{l-1}^{\bot}\mathbf{g}_j|}$.
\item[-] Set $S_l$ = $S_{l-1} \cup s_l$.
\end{itemize}
\end{algorithm}

Observing \eqref{MMSE_SIC_Cb} and \eqref{MMSE_SIC_Rs}, we propose two low-complexity algorithms of user selection to maximize $C_b\left(\mathbf{H}(S_k)\right)$ and $C_s^{\operatorname{Fk}}(\mathbf{H}(S_k),\mathbf{G}(S_k))$ , which are given at the top of this page, respectively. They are greedy-like algorithms where in each selection step the current best user is selected.

In the following, we provide a comprehensive analysis on the asymptotic ESSR of Algorithm 1 and Algorithm 2, which quantifies the multiuser secrecy gains, and reveals the achievable scaling laws of the two proposed algorithms.
With the extreme value theory, we have the following theorem.
\begin{theorem}
Let $M_l \triangleq M-l-1$, $N_l \triangleq N-l-1$ and $K_l \triangleq K-l+1$. For a given $k$, as $K \rightarrow \infty$, the ESSR of the proposed greedy algorithms for the two CSI cases in the high SNR regime can be characterized by
\begin{align}
R_{s}^{\operatorname{Mk\_gre}} & {\approx} \Bigg\{\sum_{l=1}^{k}\big[ \Psi_l+\sigma_{Ml} G_K^l\big] \Bigg\}^+, \label{Theorem_31}  \\
R_{s}^{\operatorname{Fk\_gre}} & {\approx} \Bigg\{\sum_{l=1}^{k}\Big[ \Psi_l+\sqrt{\sigma_{Ml}^2+\sigma_{Nl}^2} G_K^l\Big] \Bigg\}^+, \label{Theorem_32}
\end{align}
where $\Psi_l = \psi(M_l)-\psi(N_l)$,
$G_{K}^{l}= \sqrt{2\log K_l}-\frac{\log(4\pi \log K_l )}{2\sqrt{2 \log K_l}}-\frac{\psi{(l)}}{\sqrt{2\log K_l}}$,
$\sigma_{Ml}^2 = \frac{\pi^2}{6}-\sum_{i=1}^{M_l-1}\frac{1}{i^2},$ and $\sigma_{Nl}^2 = \frac{\pi^2}{6}-\sum_{i=1}^{N_l-1}\frac{1}{i^2}. $
\end{theorem}
%k\psi(M-k+1)+\sum_{n=1}^{k-1}\frac{n}{M-n}
%R_s^{ik^*} & \approx k \log \rho +\sum_{n=1}^{k} \log X_{(m)}^{n} - \mu_{N,k}

\begin{IEEEproof}
Please see Appendix E.
\end{IEEEproof}

%Note that the basic assumption of extreme value theory is that $K$ is sufficiency large. However, as can be shown in Fig. 3, theorem 4 behaves well even with a moderate $K$.

\textbf{\emph{Remark 3}}:  As shown in Appendix E, for the main CSI case, when the greedy user selection is adopted, the ergodic sum-rate achieved by the BS is $\mathbb{E}\big[C_b^{\operatorname{k\_opt}}(\mathbf{H})\big] = k\log\rho+\sum_{l=1}^k \left(\psi(M_l)+\sigma_{Ml} G_K^l\right)$. In contrast, the ergodic sum-rate achieved at the BS is $\mathbb{E}\big[C_b^{\operatorname{k\_ran}}(\mathbf{H})\big] = k\log\rho+\sum_{l=1}^k \psi(M_l)$ for the random user selection. On the other hand, it holds $\mathbb{E}\big[C_e^{\operatorname{k\_opt}}(\mathbf{H})\big]=
\mathbb{E}\big[C_e^{\operatorname{k\_ran}}(\mathbf{H})\big]$ for main CSI case. Hence, the multiuser secrecy gain for main CSI case is $\Delta_s^{\operatorname{Mk\_gre}} = \sum_{l=1}^k \sigma_{Ml} G_K^l$. Similarly,  the multiuser secrecy gain for the full CSI case is $\Delta_s^{\operatorname{Fk\_gre}}=\sum_{l=1}^k \sqrt{\sigma_{Ml}^2+\sigma_{Nl}^2} G_K^l$.  We can observe that the multiuser secrecy gains for the both cases depend on not only the number of total users and selected/served users but also the number of antennas at the BS and the eavesdropper. It is shown in Lemma 1, when $k\le M \le N$, both variances $\sigma_{Ml}$ and $\sigma_{Nl}$ are increasing functions of $k$, while they are decreasing functions of $M$ and/or $N$. That is, increasing $M$ and/or $N$ results in a decreasing multiuser secrecy gain, while serving more users simultaneously in each time slot can bring a larger multiuser secrecy gain to secure communications. Besides, we can note that the full CSI achieves a larger multiuser secrecy gain than the main CSI does. For the special case $M =N$, $R_{s}^{\operatorname{Fk\_opt}}$ would grow $\sqrt{2}$ times as fast as $R_{s}^{\operatorname{Mk\_opt}}$.

\textbf{\emph{Remark 4}}: Note that, in general, each user can have individual secrecy rate requirements. Let $R_s^{l}$ denote the achievable secrecy rate for the $l$-th selected user. We assume that the receiver architectures of the BS use a combination of minimum mean square estimation (MMSE) and successive interference cancellation (SIC). It is shown that the MMSE-SIC receiver achieves the capacity of the fading MIMO channel \cite[pp. 394]{ref:D. Tse}.  From Algorithm 1, we can see that the individual secrecy rate of each selected user depends on the chosen ordering, and thus the BS can use the reverse ordering of algorithm 1 for detection. However, we generally do not know the behavior of the eavesdropper, thus it is reasonable to consider a worst-case scenario where the eavesdropper can cancel multiuser interferences. Then with Lemma 1, $R_s^{l}$ is lower bounded by $\left\{\psi(M_l)+\sigma_{Ml} G_K^l-\psi(N)\right\}^+$.
\begin{figure}[t]
\begin{center}
\includegraphics[width=2.8in]{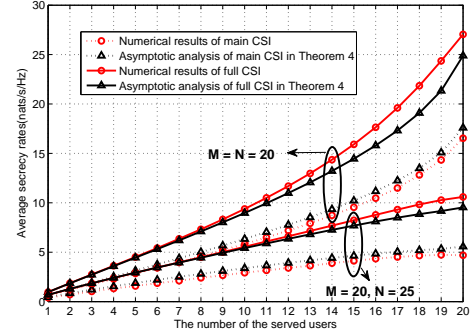}
\end{center}
\vspace{-0.3cm}
\caption{ESSR of the greedy user selection versus the number of the served users in the high SNR regime.}
\end{figure}

Fig. 3 shows the asymptotic results given in Theorem 4, and the corresponding numerical results. We set the number of total users as $K=400$ and let $\rho= 30\ dB$. It can be seen that the asymptotic results are consistent with the  numerical ones even for a moderate value of $K$.

% In the uplink, multiple-antennas always can not be equipped at the mobile users due to the limitations of size and cost. Under such a scenario, exploiting multiple user gain to improve the physical-layer security is a natural manner.

As $N>M$, since it holds that $\sum_{l=1}^{k+1}\Psi_l-\sum_{l=1}^{k}\Psi_l = -\sum_{i=M-k+1}^{N-k}\frac{1}{i}$, the first terms in \eqref{Theorem_31} and \eqref{Theorem_32} decrease as $k$ increases. Although the multiuser secrecy gains (the second terms) are increasing functions with $k$, the ESSRs may decrease with $k$ when $N$ is much larger than $M$. Therefore, an interesting observation is that serving only one user may be a favourable scheme when the eavesdropper is a more capable receiver than the BS.  In such a TDMA-like scheme, the BS serves the strongest user in each time slot to maximize $C_b(\mathbf{H},i)$ or $C_s(\mathbf{H},\mathbf{G},i)$. We have the following corollary.

\textbf{Corollary 5.} In the high SNR regime, for $k=1$, the ESSRs of the proposed greedy algorithms for the two CSI cases can be given respectively by
\begin{align}
R_{s}^{\operatorname{M1\_gre}} &\approx\left\{\psi(M)-\psi(N)+\sqrt{{2\log K}/{M}}\right\}^+, \label{Coro_51} \\
R_{s}^{\operatorname{F1\_gre}}&\approx\left\{\psi(M)-\psi(N)+\sqrt{{2\log K}\left({1}/{M}+{1}/{N}\right)}\right\}^+. \label{Coro_52}
\end{align}
\begin{IEEEproof}
Resorting to Theorem 4 and \cite[Lemma 1]{ref:B. M. Hochwald}, we can easily complete the proof.
\end{IEEEproof}

Corollary 5 shows that, when only the strongest user is served in each time slot, the multiuser secrecy gains scale with $\sqrt{\log K}$ for the both CSI cases. Besides, we further observe that the multiuser secrecy gains decrease as the number of antennas at the BS and/or the eavesdropper increase.

%Obviously, $R_{s}^{\operatorname{M1\_gre}}$ and $R_{s}^{\operatorname{F1\_gre}}$ are decreasing functions with $N$. Hence, $R_{s}^{\operatorname{M1\_gre}}$ and $R_{s}^{\operatorname{F1\_gre}}$ would converge to zero as $N$ gets large. Now we derive the value $\eta \triangleq \frac{N}{M}$ below which the ESSR is positive. For a large $M$, it holds that $\psi(M) \thickapprox \log(M-1) \thickapprox \log(M)$ \cite{ref:R. M. Young}. Therefore, when only the strongest user is served in each time slot, the existence condition for a positive ESSR is $\log \eta \le \sqrt{\frac{2\log K}{M}}$ for main CSI, while it is $\sqrt{\frac{\eta}{1+\eta}}\log \eta \le \sqrt{\frac{2\log K}{M}}$ for full CSI. Suppose that $M =20$ and $K =400$. For main CSI, the existence condition is $\eta < 2.1685$. While for full CSI, the existence condition is $\eta < 9.7480$. That is, for main CSI, a positive ESSR is still achieved even when the number of the eavesdropper antennas is twice as many as that of the BS. This can be verified in Fig. 8.
%However, there is a tradeoff for equipping the antennas of the BS.

\subsection{Low SNR Regime}
As $\rho\rightarrow 0$, with the same equation in \eqref{rho-0}, the instantaneous secrecy sum-rates with optimal $k$ user selection for the main CSI case and full CSI case can be given by
\begin{align}
C_{s}^{\operatorname{Mk\_gre}} &= \left\{\Big[\underset{S_{k}}{\max} \ \ {C_b^k\left(\mathbf{H}(S_k)\right)}\Big]-C_e^k(\mathbf{G}(S_k))\right\}^+ \nonumber \\
&\approx \rho\left\{\Big[\underset{S_{k}}{\max} \ X(\mathbf{H}(S_k))\Big]-Y(\mathbf{G}(S_k))\right\}^+,  \label{Cs_opti_Low} \\
C_{s}^{\operatorname{Fk\_gre}} &= \Big\{\underset{S_{k}}{\max} \ \ \Big[{C_b^k\left(\mathbf{H}(S_k)\right)}-C_e^k(\mathbf{G}(S_k)) \Big] \Big\}^+  \nonumber \\
&\approx \rho\left\{\underset{S_{k}}{\max} \Big[ X(\mathbf{H}(S_k))-Y(\mathbf{G}(S_k))\Big]\right\}^+, \label{Cs_optii_Low}
\end{align}
where $X(\mathbf{H}(S_k)) = \operatorname{tr}(\mathbf{H}(S_k)^{\dagger} \mathbf{H}(S_k))$ and $Y(\mathbf{G}(S_k)) = \operatorname{tr}(\mathbf{G}(S_k)^{\dagger} \mathbf{G}(S_k))$.

According to Equations \eqref{Cs_opti_Low} and \eqref{Cs_optii_Low}, the optimal user selection in the low SNR regime is to maximize $X(\mathbf{H}(S_k))$ or $X(\mathbf{H}(S_k))-Y(\mathbf{G}(S_k))$. The key idea behind this is to select $k$ users with the strongest channels, thus the optimal user selection is equivalent to the greedy one in the low SNR regime. Let $\Xi$ denote the sequence with variables $\Xi_i = \mathbf{h}_i^\dag \mathbf{h}_i, i=1,\cdots,K$, then the instantaneous secrecy sum-rate given in \eqref{Cs_opti_Low} can be rewritten as
\begin{align}
C_{s}^{\operatorname{Mk\_gre}} &\approx \rho\left\{\sum_{r=1}^{k} \Xi_{(r)}^K-\operatorname{tr}(\mathbf{G_*}^{\dagger} \mathbf{G_*})\right\}^+,
\end{align}
where $\mathbf{G_*}$ is the channel between the $k$ selected users and the eavesdropper. Let $\Psi$  denote the sequence with variables $\Psi_i = \mathbf{h}_i^\dag \mathbf{h}_i - \mathbf{g}_i^\dag \mathbf{g}_i, i=1,\cdots,K$, then the instantaneous secrecy rate given in \eqref{Cs_optii_Low} can be rewritten as
\begin{align}
C_{s}^{\operatorname{Fk\_gre}} &\approx \rho\left\{\sum_{r=1}^{k} \Psi_{(r)}^K \right\}^+.
\end{align}

Since the entries of $\Xi$ are $\chi_{2M}^2$ variables, conventionally, the Gumbel distribution can be used to approximate the asymptotic distribution of $\Xi_{(1)}^K$ \cite{ref:D. Bai}. It holds
\begin{align}
\underset{K\rightarrow \infty}{\lim}\mathbb{P}\left(\Xi_{(1)}^K\le c_K t+d_K\right)=e^{-e^{-t}},
\end{align}
where $c_K=1, d_K=\log K+(M-1)\log \log K-\log \Gamma(M)$.

The above asymptotic distribution of $\Xi_{(1)}^K$ has been widely used in the existing works \cite{ref:A. Goldsmith}, \cite{ref:M. Sharif Partial}. However, for i.i.d. Chi-square random variables, the convergence of its maximum to the Gumbel distribution with parameters $c_K$ and $d_K$ is quite slow \cite{ref:J. Kampeas}. Specially, for a large $M$, $\Gamma(M)$ is an excessively large number, i.e., $\Gamma(20)=1.2165\times 10^{17}$, thus $d_K$ may not be  positive for a moderate value of $K$. Therefore, we calculate new normalizing constants of $\Xi_{(1)}^K$ by approximating Chi-square random variables as Gaussian ones. The reasons are: i) as shown in Theorem 2 and Fig. 2, a Chi-square random variable can be well approximated by a Gaussian one \cite{ref:D. Horgan}; ii) for i.i.d. Gaussian random variables, the convergence of the maximum to the Gumbel distribution using the normalizing constants given in \eqref{MAX_con_a} and \eqref{MAX_con_b} behaves well even for a not so large $K$ \cite[pp. 302]{ref:David}. Note that we focus on the expectation of the maximum, which helps to provide a closed-form expression for the ESSR. In order to verify the accuracy of the analytical expectation with Gaussian approximation, some numerical results are given in Table I, where we let $K=100$. It is shown that the analytical results of Gaussian approximation are very close to the numerical results for all the cases, while the analytical results given in \cite{ref:A. Goldsmith} and \cite{ref:M. Sharif Partial} are not consistent with the numerical ones as $M$ gets large.

\begin{table}[]
\begin{center}
\caption{The comparison between the numerical and the analytical results of $\mathbb{E}\left[{\Xi_{(1)}^K}\right]$.}
\begin{tabular}{|c|c|c|c|}
\hline
Degrees of freedom & Numerical& Analytical & Ref. [12, 17] \\
\hline
\hline
$M = 10$ & 19.7119 & 18.0842 & 18.3498\\
\hline
$M = 20$ & 33.0152& 31.4328 & 33.6216\\
\hline
$M = 30$ &  45.5815& 44.0023 & 48.8934\\
\hline
$M = 40$ &  57.6518& 56.1684 &64.1652\\
\hline
$M = 50$ &  69.5151& 68.0768 &79.4370\\
\hline
\end{tabular}
\end{center}
\end{table}

Therefore, the entries of $\Xi$ and $\Psi$ are treated as Gaussian random variables. Then with Lemma 3, we can present the following theorem.
\begin{theorem}
In the low SNR regime, for a given $k$, the ESSRs of the greedy $k$ users selection in the low SNR regime are given by
\begin{align}
R_{s}^{\operatorname{Mk\_gre}} &\approx k \rho\Big\{R_{b}^{\operatorname{Mk\_gre}}-N\Big\}^+, \\
R_{s}^{\operatorname{Fk\_gre}} &\approx k \rho\Big\{R_{b}^{\operatorname{Fk\_gre}}-N\Big\}^+,
\end{align}
where $R_{b}^{\operatorname{Mk\_gre}}=M+\sqrt{{2M\log K}}-\frac{\log(4\pi \log K)}{2\sqrt{2\log K/M}}-\rho{\sqrt\frac{M}{2\log K}}$
$\big(\psi(k+1)-1\big)$ and $R_{b}^{\operatorname{Fk\_gre}}=M+\sqrt{{2(M+N)\log K}}-\frac{\log(4\pi \log K)}{2\sqrt{2\log K/(M+N)}} {\sqrt\frac{M+N}{2\log K}-(\psi(k+1)-1)}$.
\end{theorem}

\begin{IEEEproof}
Invoking $\sum_{r=1}^{k}\psi(r)/k=\psi(k+1)-1$, along with the results given in Lemmas 2 and 3, we can easily complete the proof.
\end{IEEEproof}
\begin{figure}[t]
\begin{center}
\includegraphics[width=3in]{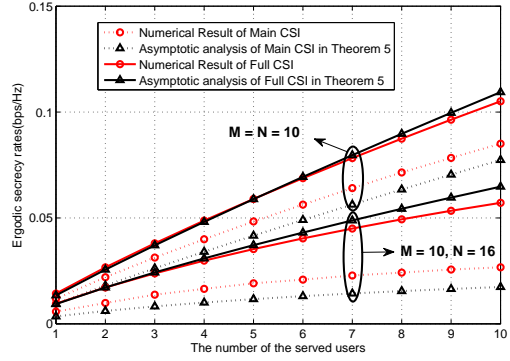}
\end{center}
\vspace{-0.3cm}
\caption{ESSR of the greedy user selection versus the number of the served users in the low SNR regime.}
\end{figure}

\textbf{\emph{Remark 5}}: Differing from the high SNR regime, the multiuser interference is negligible in the low SNR regime. Hence, the individual ergodic secrecy rate of the $l$-th selected users equals to that of the case with an interference-free eavesdropper in the low SNR regime, which can be approximated as $R_s^{l}\approx \rho\Big\{R_{b}^{\operatorname{Mk\_gre}}-N\Big\}^+$.

In Fig. 4, we employ simulations to verify Theorem 5. We set $K=400$ and let $\rho= -30 \ dB$. Again, it can be seen that the asymptotic results can describe the behaviour of the numerical results well for the both CSI cases.
%% and ignoring the correlation between ,

Similar to Corollary 5, we present the following corollary to characterise the multiuser secrecy gain of TDMA-like scheme in the low SNR regime.

\textbf{Corollary 6.}  In the low SNR regime, for $k=1$, the ESSRs of the proposed greedy algorithms for the two CSI cases can be given respectively by
\begin{align}
R_{s}^{\operatorname{M1\_gre}} &\approx\rho\left\{M-N+\sqrt{{2{M}\log K}}\right\}^+, \label{Coro_61} \\
R_{s}^{\operatorname{F1\_gre}}&\approx\rho\left\{M-N+\sqrt{{2(M+N)\log K}}\right\}^+. \label{Coro_62}
\end{align}

%Especially, as $M=N$, the ESSR for $k=1$ can be given by
%\begin{align}
%R_{s}^{\operatorname{M1\_gre}} &\approx \rho\sqrt{{2{M}\log K}}, \label{Coro_63} \\
%R_{s}^{\operatorname{F1\_gre}} &\approx \rho \sqrt{{4{M}\log K}}. \label{Coro_64}
%\end{align}
\begin{IEEEproof}
Resorting to Theorem 5 and \cite[Lemma 1]{ref:B. M. Hochwald}, we can easily obtain the results.
\end{IEEEproof}

Unlike the high SNR case, the multiuser secrecy gains grow with $M$ and $N$ for the low SNR case. We can also note that, as $N$ increases, the achievable rate at the eavesdropper grows logarithmically with $N$ in the high SNR regime, while it increases linearly with $N$ in the low SNR regime.

%In the low SNR regime, the existence condition of a positive ESSR for main CSI is $\eta < 1+ \sqrt{\frac{2\log K}{M}}$. While it is $\frac{\eta-1}{\sqrt{\eta+1}} < \sqrt{\frac{2\log K}{M}}$ for full CSI. Again, we consider the case $M=20$ and $K=400$. The existence conditions for the both CSI cases are $\eta < 1.7740$ and  $\eta < 2.4340$, respectively. Let us compare the numerical results of existence condition between the low and high SNR regimes, we can observe that the multiuser secrecy gain helps to tolerate a more capable eavesdropper in the high SNR regime than that in the low SNR regime. This can be demonstrated by Corollary 5 and 6.

\subsection{Large Scale Analysis for Greedy User Selection}
By using a very large antenna array, the achievable rate of each user in a multiuser MIMO system is equal to that of a single-input multiple-output (SIMO) system, without any inter-user interference \cite{ref:H. Q. Ngo}. In the following, we use this potential for user selection, and derive asymptotic results for the ESSR. The interesting operating regime is when both $M$ and $N$ are enough large, and they are much larger than $k$, i.e., $k \ll M$ and $k \ll N$. In a large scale multiple antenna system, according to \cite[Theorem 1]{ref:B. M. Hochwald}, the approximated distribution of the mutual information, $C_b^k(\mathbf{H})$, can be given as
\begin{align}
C_b^k(\mathbf{H}) \sim \mathcal{N}\left(k\log\left(1+\rho M\right),\frac{k}{M}\right).
\end{align}

Obviously, it holds $C_b^1(\mathbf{H}) \sim \mathcal{N}\left(\log\left(1+\rho M\right),\frac{1}{M}\right)$ as $k=1$. As $M$ grows large, the variances $\frac{k}{M}$ and $\frac{1}{M}$ converge to zero. We can conclude that, with large $M$, both $C_b^k(\mathbf{H})$ and $C_b^1(\mathbf{H})$ would converge to their mean, and it holds $C_b^k(\mathbf{H}) = k C_b^1(\mathbf{H})$. This observation further verifies that the inter-user interference vanishes in the large scale antenna systems.

Since the inter-user interference is negligible, the basic idea of user selection is same as that in the low SNR regime, where the $k$ users with the strongest channels are selected to communication with. Note that this scheme, referred to as norm based greedy selection, only requires each user to calculates the squared Frobenius norm of its wireless channel and can be implemented in a distributed manner \cite{ref:Bletsas}, it is especially attractive in large scale systems. We give the asymptotic results of the ESSR for large scale systems in the following Theorem.

\begin{theorem}
In a large scale system, for a fixed $k$, the ESSRs of the greedy user selection are approximately given by
\begin{align}
R_{s,lar}^{\operatorname{Mk\_gre}} \approx &\Bigg\{k\log\frac{1+\rho M}{1+\rho N}+\sum_{r=1}^{k}\sqrt{\frac{1}{M}}\left(\sqrt{2\log K}  \right. \nonumber \\
&\left.\ \ -\frac{\log(4\pi \log K)+2\psi(r)}{2\sqrt{2 \log K}}\right)\Bigg\}^+, \\
R_{s,lar}^{\operatorname{Fk\_gre}} \approx &\Bigg\{k\log\frac{1+\rho M}{1+\rho N}+\sum_{r=1}^{k}\sqrt{\frac{1}{M}+\frac{1}{N}}\left(\sqrt{2\log K}\right.  \nonumber \\
&\left.\ \ -\frac{\log(4\pi \log K)+2\psi(r)}{2\sqrt{2 \log K}}\right)\Bigg\}^+.
\end{align}
\end{theorem}

\begin{IEEEproof}
Resorting to Lemma 3 and \cite[Theorem 1]{ref:B. M. Hochwald}, we can easily prove Theorem 6.
\end{IEEEproof}

\begin{figure}[t]
\begin{center}
\includegraphics[width=3in]{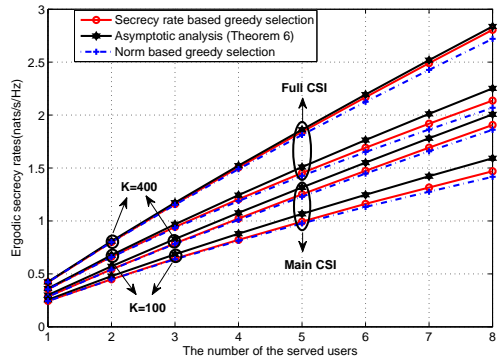}
\end{center}
\vspace{-0.3cm}
\caption{The ESSR of greedy user selection versus the number of selected users in large scale systems.}
\end{figure}

In Fig. 5, we depict the numerical and asymptotic results of the proposed user selection scheme for large scale systems. The numerical results of the secrecy rate based user selection scheme are also plotted. We set the parameters as $M=N=100$ and $\rho=10\ dB$. We can see that, the norm based greedy selection achieves a ESSR near to that of the secrecy rate based scheme. This further confirms that the norm based user selection is a good choice in the large scale system. As expected, the asymptotic results agree well on the numerical ones. We can also note that, although channel hardening happens when the BS and the eavesdropper equip with a large number of antennas, the multiuser gain still brings a significant secrecy improvement to wireless communications.
\section{Effect of Channel Estimation Errors}
So far, we assume that the main CSI or full CSI are perfectly known at the BS. However, the BS can only obtain a noisy version of the CSIs by channel estimation in practice. In this section, we investigate the impact of channel estimation errors on the performance of user selection in secure communication. Suppose that the BS uses the minimum mean square error (MMSE) estimator, the $i$-th user's channel gain can be modeled by
\begin{align}
\mathbf{h}_i = \sqrt{1-\xi}\hat{{\mathbf{h}}}_i + \sqrt{\xi}\tilde{{\mathbf{h}}}_i,
\end{align}
where $\sqrt{1-\xi}\hat{{\mathbf{h}}}_i$ is the estimation of the $i$-th user's channel gain, $\sqrt{\xi}\tilde{{\mathbf{h}}}_i$ is the estimation error, and $\xi$ is the error variance $(\xi \in (0,1))$. The channel vector $\hat{{\mathbf{h}}}_i$ and $\tilde{{\mathbf{h}}}_i$ follow with $\mathcal{CN}(\mathbf{0},\mathbf{I})$. Under channel estimation errors, a lower bound for the achieved sum-rate at the BS is \cite{ref:Yoo}
\begin{align}
\hat{C}_b^k(\mathbf{H})&\ge \hat{C}_{b,low}^k =\log\left|\mathbf{I}+\frac{(1-\xi)\rho}{1+\xi\rho}\hat{\mathbf{H}}^{\dagger} \hat{\mathbf{H}}  \right|,  \label{Rate_BS_Err}
\end{align}
where $\hat{\mathbf{H}}=[\hat{{\mathbf{h}}}_1,\cdots,\hat{{\mathbf{h}}}_k]$. It has been shown that the lower bound is tight in \cite{ref:Yoo}. Comparing \eqref{Rate_BS_Err} to \eqref{Rate_BS_Per}, we observe that the channel estimation errors result in a SNR loss factor of at most $\eta\triangleq\frac{1-\xi}{1+\xi\rho}$. Here, we consider the worst-case that the eavesdropper has the perfect CSI. Therefore, the secrecy sum-rate for an uplink transmission is lower bounded by
\begin{align}
\hat{C}_s^k \ge \left\{\log\left|\mathbf{I}+\frac{(1-\xi)\rho}{1+\xi\rho}\hat{\mathbf{H}}^{\dagger} \hat{\mathbf{H}}  \right|-\log\left|\mathbf{I}+\rho\mathbf{G}^{\dagger} \mathbf{G}  \right|\right\}^+. \label{ISR_err}
\end{align}

Next, we consider the impact of channel estimation errors on the performance of random user selection and greedy user selection, respectively.
\subsection{Random User Selection}
From \cite[Theorem 3]{ref:Z. Wang}, the variable $\hat{C}_{b,low}^k$ is also approximately a Gaussian variable with mean $\mathbb{E}[\hat{C}_{b,low}^k]$ and variance $\mathbb{V}[\hat{C}_{b,low}^k]$. The calculations of $\mathbb{E}[\hat{C}_{b,low}^k]$ and $\mathbb{V}[\hat{C}_{b,low}^k]$ are given in \cite[Theorem 3]{ref:Z. Wang}. Then with Theorem 1, we can obtain the lower bound for $\hat{R}_{s}^ {\operatorname{k\_ran}}$ under channel estimation errors. Although a closed-form expression of this lower bound for all SNR regimes is not trivial, explicit results for the low SNR case can be given.
In the low SNR regime, it holds $\eta=\frac{1-\xi}{1+\xi\rho}\thickapprox 1 - \xi$. Therefore, we have
\begin{align}
\hat{R}_{s}^ {\operatorname{k\_ran}} &\gtrapprox \mathbb{E}\left[\left\{\log\left|\mathbf{I}+(1-\xi)\rho \hat{\mathbf{H}}^{\dagger} \hat{\mathbf{H}}  \right|-\log\left|\mathbf{I}+\rho\mathbf{G}^{\dagger} \mathbf{G}  \right|\right\}^+\right] \nonumber \\
&\thickapprox\frac{\hat{\beta}_k\rho}{\sqrt{2\pi}}e^{-\frac{\hat{\alpha}_k^2}{2\hat{\beta}_k^2}}
+\frac{\hat{\alpha}_k\rho}{2}\left(1+\operatorname{erf}\left(\frac{\hat{\alpha}_k}{\sqrt{2} \hat{\beta}_k}\right)\right),
\end{align}
where $\hat{\alpha}_k = k\left((1-\xi)M-N\right)$, and  $\hat{\beta}_k^2 = k\left((1-\xi)^2M+N\right)$.

Let us consider a special case $\hat{\alpha}_k=0$, it holds $M=\frac{N}{1-\xi}$ and $\hat{R}_{s}^ {\operatorname{k\_ran}} \gtrapprox \rho\sqrt \frac{k(2-\xi)N}{2\pi}$. In contrast, when $M=N$, it holds ${R}_{s}^ {\operatorname{k\_ran}} \approx \rho\sqrt \frac{kN}{\pi}$ under perfect channel estimation.

%We can observe that the ESSR achieved with more BS antennas in the imperfect CSI case is still lower than that in the perfect CSI case.
\subsection{Greedy User Selection}
For greedy user selection, we only focus on the main CSI case. Under channel estimation errors, the user selection problem is to select a set of channel estimations $\hat{\mathbf{H}}(S_k)=[\hat{\mathbf{h}}_{s_1},\hat{\mathbf{h}}_{s_2},\cdots,\hat{\mathbf{h}}_{s_k}]$ such that $\hat{C}_{b,low}^k$ is maximized. Similar to the procedure of \eqref{MMSE_SIC_Cb}, we have
\begin{align}
\hat{C}_{b,low}^k =  \sum_{l=1}^{k}\log\left|1+\xi\rho \hat{\mathbf{h}}_{s_l}^{\dagger}\hat{\mathbf{A}}_{l-1}^{\bot} \hat{\mathbf{h}}_{s_l} \right|,
\end{align}
where $\hat{\mathbf{A}}_{l-1}^{\bot} =\mathbf{I}_M - \xi\rho \hat{\mathbf{H}}(S_{l-1})\Big(\mathbf{I}_{l-1}+\xi\rho\hat{\mathbf{H}}(S_{l-1})^{\dagger}$
$\hat{\mathbf{H}}(S_{l-1})\Big)^{-1}\hat{\mathbf{H}}(S_{l-1})^{\dagger}$.
 Therefore, the greedy user selection under channel estimation errors can also be implemented in the way as Algorithm 1. We do not present the explicit algorithm here due to space limitations. Similar to the random user selection, a closed-form lower bound of ESSR under channel estimation errors in the low SNR regime is given as
 \begin{align}
 &\hat{R}_{s}^{\operatorname{Mk\_gre}} \gtrapprox \Bigg\{k \rho\left((1-\xi)\left(\sqrt{{2M\log K}}-\frac{\log(4\pi \log K)}{2\sqrt{2\log K/M}}\right.\right. \nonumber \\
 &\left. \left.+M\right)-N\right)-{\sqrt\frac{\rho^2(1-\xi)^2 M}{2\log K}}\sum_{r=1}^{k}\psi(r)\Bigg\}^+.
 \end{align}

For a meaningful comparison, let us consider the TDMA-like scheme, where only one user is served in each time slot. Through steps of mathematical manipulations, we further have
\begin{align}
\hat{R}_{s}^{\operatorname{M1\_gre}} \gtrapprox \rho\left\{(1-\xi)M-N+(1-\xi)\sqrt{{2{M}\log K}}\right\}^+.  \label{Coro_61_Err}
\end{align}

Comparing \eqref{Coro_61_Err} to \eqref{Coro_61}, we can observe that the TDMA-like scheme under channel estimation errors not only achieves a lower ESSR, but also has less benefit from user selection, than that under perfect channel estimation.
\section{Simulation Results}
In this section, we further investigate the secrecy performance of the proposed schemes numerically. We perform Monte Carlo experiments each with 10000 independent trials to obtain the numerical results.

\begin{figure}[t]
\begin{center}
\includegraphics[width=2.8in]{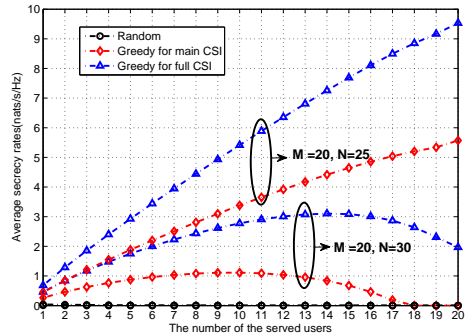}
\end{center}
\vspace{-0.3cm}
\caption{Comparisons between the random and greedy user selection in the high SNR regime.}
\end{figure}

Fig. 6 demonstrates the secrecy gain achieved by the greedy user selection in Algorithms 1 and 2. As can be seen, the ESSR of the random user selection scheme almost converges to zero for all the cases. In contrast, the greedy user selection scheme provides a much higher ESSR even when the eavesdropper has five more antennas than the BS. This verifies that the secrecy gain achieved from user selection is fairly prominent. We also note that as the number of eavesdropper antennas increases, the optimal number of served users becomes small, which will be further studied in Fig. 9.

\begin{figure}[t]
\begin{center}
\includegraphics[width=2.8in]{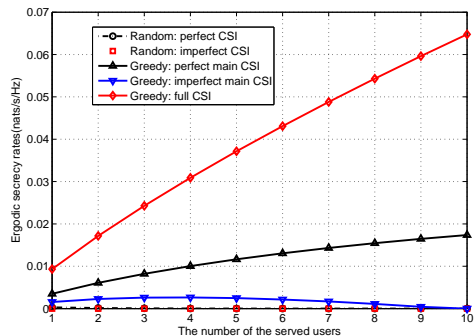}
\end{center}
\vspace{-0.3cm}
\caption{Comparisons between the random and greedy user selection in the low SNR regime.}
\end{figure}

\begin{figure}[t]
\begin{center}
\includegraphics[width=2.8in]{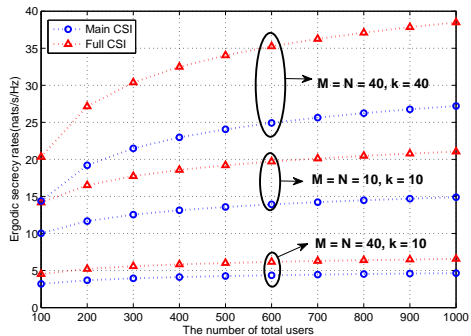}
\end{center}
\vspace{-0.3cm}
\caption{The ESSR of greedy user selection versus the number of total users.}
\end{figure}

Similar to the high SNR case, the greedy user selection also achieves a significant secrecy gain in the low SNR regime, which is demonstrated in Fig. 7, where $\rho= -30\ dB$, $M=10$, $N=15$, and $\xi=0.1$. We can note that, when the greedy user selection is adopted, the ESSR grows with $k$ even if the eavesdropper has five more antennas than the BS under perfect CSI. As expected, channel estimation errors result in a large ESSR loss. However, the greedy user selection under imperfect CSI still achieves secrecy multiuser gains.

Fig. 8  plots the ESSRs versus the number of total users for different numbers of served users and antennas. The SNR is fixed to $30\ dB$ for all curves. We can observe that all the ESSRs increase as $K$ increases, because a high multiuser secrecy gain can be achieved. Let us compare the case $M=N=10, k =10$ with the one $M=N=40, k =10$, we note that the multiuser secrecy gain decreases for larger $M$ and $N$. However, the multiuser secrecy gain grows with $k$. This is because as $k$ gets large, $\mathbf{H}^{\dagger} \mathbf{H}$ and $\mathbf{G}^{\dagger} \mathbf{G}$ do not converge to a deterministic quantity and thus the tail probability becomes large. Therefore, a significant multiuser selection gain exists. The simulations in the low SNR regime look similar as those in the high SNR case and thus are omitted.

\begin{figure}[t]
\begin{center}
\includegraphics[width=2.8in]{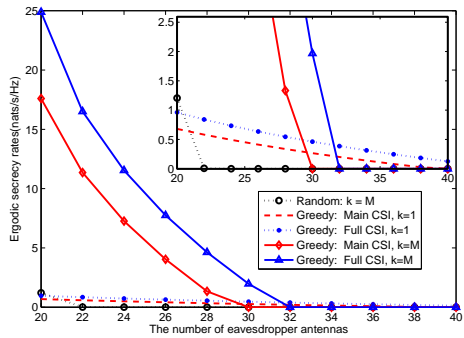}
\end{center}
\vspace{-0.3cm}
\caption{The ESSR of the proposed greedy schemes versus the number of eavesdropper antennas.}
\end{figure}

In Fig. 9, we investigate the impact of the number of the eavesdropper antennas on secrecy performance, where $M=20$ and $K = 400$. Since the random selection scheme offers no multiuser secrecy gain, it can not provide a positive ESSR even when the eavesdropper equips two more antennas than the BS. Although serving up to $M$ users approaches the maximal multiplexing gain, the secrecy performance of the greedy user selection scheme for both main CSI and full CSI deteriorates quickly as the number of eavesdropper antennas gets large. Hence, we can conclude that only serving the strongest user in each time slot is a favorable scheme when there is a more capable eavesdropper than the BS in the network.

%The results for the case with a larger $\rho$ look similar as Fig. 2 and are omitted due to space limitations.

\section{Conclusion}
In this paper, we have investigated the problem of user selection in a multiuser uplink communication system with a multiple antennas eavesdropper. Closed-form expressions of the ESSR for random user selection in the both low and high SNR regimes have been derived. We have shown that when the eavesdropper has the same number of antennas with the BS, the maximum ESSR scales with $\sqrt{\log M}$ in the high SNR regime, while scales with $M$ in the low SNR regime. To achieve the multiuser secrecy gain, we have proposed two greedy user selection algorithms for main CSI and Full CSI, respectively. The corresponding closed-form expressions of the ESSR have also been presented. We have revealed that the multiuser secrecy gain not only depends on the number of total users, but also depends on the number of served users, the BS and the eavesdropper antennas. Furthermore, the impact of the number of the eavesdropper antennas and channel estimation errors on the ESSR performance have been explored.

\appendices
\section{Proof of Lemma 2}
When $\{\mathbf{Y}_K\}$ is a sequence of i.i.d. $\mathcal{N}(\mu,\sigma^2)$ variables, $\frac{Y_i-\mu}{\sigma}$ is distributed according to $\mathcal{N}(0,1)$ for $i \in \{1,2,\cdots,K\}$. If $Y_{(1)}^K$ is the maximum of the sequence $\{\mathbf{Y}_K\}$,  $\frac{Y_{(1)}^K-\mu}{\sigma}$ is therefore the maximum of the sequence $\left\{\frac{Y_{K}-\mu}{\sigma}\right\}$ which satisfies \cite[pp. 302]{ref:David}
\begin{align}
  \underset{K\rightarrow \infty}{\lim}\mathbb{P}\left(\frac{Y_{(1)}^K-\mu}{\sigma}\le a_K{'} t+b_K{'}\right)=e^{-e^{-t}},\label{Gumbel}
\end{align}
where $a_K{'}=\frac{1}{\sqrt{2\log K}}$ and $b_K{'}=\sqrt{2\log K} - \frac{\log(4\pi \log K)}{2\sqrt{2 \log K}}$. We further rewritten \eqref{Gumbel} as
\begin{align}
\underset{K\rightarrow \infty}{\lim}\mathbb{P}\left(Y_{(1)}^K\le a_K t+b_K\right)=e^{-e^{-t}},
\end{align}
where $a_K=\frac{\sigma}{\sqrt{2\log K}}$,  and $b_K=\sigma\sqrt{2\log K} -\sigma \frac{\log(4\pi \log K)}{2\sqrt{2 \log K}}+\mu$. Since $G(t)=e^{-e^{-t}}$ is the Gumbel distribution, as $K$ increases, the mean of $Y_{(1)}^K$ approaches %[Capacity of MIMO Channels With Antenna Selection]
\begin{align}
\mathbb{E}[Y_{(1)}^K]&\overset{(a)}{=}a_K\gamma+b_K  \nonumber \\
&=\sigma\sqrt{2\log K} -\sigma \frac{\log(4\pi \log K)-2\gamma}{2\sqrt{2 \log K}}+\mu,    \label{MAX_mean}
\end{align}
where step $(a)$ follows from the result given in \cite[pp. 298]{ref:David}. We complete the proof.

\section{Proof of Lemma 3}
%For a $\mathcal{N}(\mu,\sigma^2)$ sequence, the $r$-th largest extreme holds that \cite{ref:J. Galambos}
%\begin{align}
%\underset{K\rightarrow \infty}{\lim}\mathbb{P}\left(Y_{(r)}^K\le a_K t+b_K\right)=e^{-e^{-t}}\sum_{i=0}^{r-1}\frac{e^{-ti}}{\Gamma(i+1)},
%\end{align}
%where $a_K$ and $b_K$ are given in \eqref{MAX_con_a} and \eqref{MAX_con_b}, respectively.
When $\{\mathbf{Y}_K\}$ is a sequence of i.i.d. $\mathcal{N}(\mu,\sigma^2)$ variables, $\frac{Y_i-\mu}{\sigma}$ obeys $\mathcal{N}(0,1)$ for $i \in \{1,2,\cdots,K\}$. If $Y_{(r)}^K$ is the $r$-th largest order statistic of the sequence $\{\mathbf{Y}_K\}$, $\frac{Y_{(r)}^K-\mu}{\sigma}$ is therefore the $r$-th largest order statistic of the sequence $\left\{\frac{Y_{K}-\mu}{\sigma}\right\}$. Let $Z_{(r)}^K \triangleq \frac{Y_{(r)}^K-\mu}{\sigma}$, which satisfies \cite{ref:R. L. Smith}
\begin{align}
\mathbb{E}\left[\left(Z_{(r)}^K\right)^s e^{-\zeta Z_{(r)}^K}\right]=(-1)^{s}\frac{\Gamma^{(s)}(r)}{\Gamma{(r)}}.  \label{Smith_mean}
\end{align}

The mean of $Z_{(r)}^K$ can be given by setting $\zeta=0$ and $s=1$ in \eqref{Smith_mean}, i.e.,
\begin{align}
\mathbb{E} \left[Z_{(r)}^K\right] = \frac{-\Gamma^{(1)}(r)}{\Gamma{(r)}}=-\psi(r),
\end{align}

from which we directly obtain the mean of $Y_{(r)}^K$, given by
\begin{align}
\mathbb{E} \left[Y_{(r)}^K\right] = b_K - a_K \psi(r).
\end{align}

We complete the proof.

\section{Proof of Corollary 2}
It holds $\mu_k = 0$ for the case $M =N$, we can obtain the following approximation from Theorem 1,
\begin{align}
R_s^{k} \approx \frac{\sigma_k}{\sqrt{2\pi}}.
\end{align}

Obviously, $R_s^{k}$ grows with $\sigma_k$. Since the expressions of $\sigma_k$ are different for the cases $k \le M$ and $k >M$, $R_s^k$ can be rewritten as
\begin{align}
R_s^k\approx\left\{ \begin{aligned}
&\sqrt{\frac{1}{\pi}\left(\sum_{i=1}^{k-1}\frac{i}{(M-k+i)^2}+\frac{k}{M}\right)},\ \ k \le M,  \nonumber \\
&\sqrt{\frac{1}{\pi}\left(\sum_{i=1}^{M-1}\frac{i}{(k-M+i)^2}+\frac{M}{k}\right)}, \ \ k > M.
\end{aligned} \right. \nonumber
\end{align}

Let $\varrho(k)=\sum_{i=1}^{k-1}\frac{i}{(M-k+i)^2}+\frac{k}{M}$ and $\varsigma(k)=\sum_{i=1}^{M-1}\frac{i}{(k-M+i)^2}+\frac{M}{k}$. Since it holds that
\begin{align}
&\varrho(k+1)-\varrho(k) \nonumber \\
=&\frac{k}{(M-1)^2}+\sum_{i=1}^{k-1}\frac{i\left(2(M-k+i)-1\right)}{(M-k-1+i)^2(M-k+i)^2} +\frac{1}{M} \nonumber \\
>& 0, \nonumber
\end{align}
\begin{align}
&\varsigma(k+1)-\varsigma(k) \nonumber \\
=&\sum_{i=1}^{M-1}\left[\frac{i}{(k+1-M+i)^2}-\frac{i}{(k-M+i)^2}\right]-\frac{M}{k(k+1)} \nonumber \\
< &0, \nonumber
\end{align}
$R_s^k$ is a monotonically increasing function of $k$ as $k \le M$, whereas is a monotonically decreasing function of $k$ as $k \ge M$. The maximum ESSR is clearly achieved at $k= M$, which is
\begin{align}
R_s^M &\approx \sqrt{\frac{1}{\pi}\left(\psi(M)+\gamma+1\right)} \nonumber \\
&\overset{a} \approx \sqrt{\frac{1}{\pi}\Big(\log (M-1) +\gamma + 1\Big)},
\end{align}
where step $a$ holds for a large $M$ \cite{ref:R. M. Young}. By now, the proof is completed.

\section{Proof of Theorem 4}
Let us first prove \eqref{Theorem_31}. For the \emph{main CSI} case, the BS selects $k$ users to maximize the sum-rate $C_b^k\left(\mathbf{H}(S_k)\right)$. Let $X_{lj} = \mathbf{h}_j^{\dagger}\mathbf{A}_{l-1}^{\bot}\mathbf{h}_j$ and $Y_{lj} = \mathbf{h}_j^{\dagger}\mathbf{B}_{l-1}^{\bot}\mathbf{h}_j$. Since $\mathbf{A}_{l-1}^{\bot}$ is a complex Wishart distribution with $M-l+1$ degrees of freedom, $X_{lj} $ obeys $\chi_{2(M-l+1)}^2$ \cite{ref:Mathew}. Similarly, the variable $Y_{lj}$ obeys $\chi_{2(N-l+1)}^2$. From \eqref{MMSE_SIC_Cb}, we have
\begin{align}
\max C_b\left(\mathbf{H}(S_k)\right) \approx \sum_{l=1}^{k} \underset{1 \le j \le K-l+1} {\max} \log X_{lj}.
\end{align}
%That is, the sequence of $\underset{1 \le l \le K-l+1} {\max} \log X_{lj}$ are dependent.

Based on Lemma 1, we know that the random variable $\log X_{lj}$ can be approximated by a Gaussian $\mathcal{N}\big(\psi{(M-l+1)},\sigma_{M-l+1}^2\big)$. Note that the first user is selected out of the $K$ users, there might be mild dependence between the channel of the selected user and that of the remaining users \footnote{In most works \cite{ref:A. Goldsmith}, \cite{ref:M. Sharif Partial} and \cite{ref:M. A. Maddah-Ali}, they ignore the dependence between the channel of the selected users and that of the remain users, and thus the results therein are actually upper bounds for the exact throughput. Although the authors  in \cite{ref:G. Dimic} have considered such dependence, the result is also an upper bound. In \cite{ref:S. Ozyurt}, the authors have obtained a closed-form expression of the exact joint PDF, which unfortunately are too intractable to analyze the multiuser gain.}. For analytical tractability, we assume that the sequence is i.i.d., and thus Lemma 3 can be applied, which yields
\begin{align}
\mathbb{E}\Big[{\underset{1 \le j \le K-l+1} {\max} \log X_{lj}} \Big]= \psi{(M_l)} + \sigma_{Ml}G_{K}^{l},
\end{align}
where $G_{K}^{l} = \sqrt{2\log K_l}-\frac{\log(4\pi \log K_l )}{2\sqrt{2 \log K_l}}-\frac{\psi{(l)}}{\sqrt{2\log K_l}}$,
$\sigma_{Ml}^2 = \frac{\pi^2}{6}-\sum_{i=1}^{M_l-1}\frac{1}{i^2}$,
$M_l = M - l +1$, and $K_l  = K-l+1$. Accordingly, the ESSR for main CSI case can be given by
\begin{align}
R_{s}^{\operatorname{Mk\_gre}}&\overset{(a)}=\mathbb{E}\left\{\max C_b^k\left(\mathbf{H}(S_k)\right)-C_e^k\left(\mathbf{G}(S_k)\right)\right\}^+ \nonumber \\
& \overset{(b)}{\gtrapprox} \Big\{\mathbb{E}\left[\max C_b^k\left(\mathbf{H}(S_k)\right)\right]-\mathbb{E}\left[C_e^k\left(\mathbf{G}(S_k)\right)\right]\Big\}^+ \nonumber \\
&\approx  \Bigg\{\sum_{l=1}^{k}\big[ \psi(M_l)-\psi(N_l)+\sigma_{Ml} G_K^l\big] \Bigg\}^+, \label{Rs_ikopt}
\end{align}
where $N_l = N-l+1$ and the result in $(b)$ is obtained by applying Jensen's inequality to $(a)$. As $K$ in $(b)$ goes to infinity, we obtain the result given in \eqref{Theorem_31}.

Next we prove the result given in \eqref{Theorem_32} by starting from the secrecy sum-rate expression for \emph{full CSI} given in \eqref{MMSE_SIC_Rs}
\begin{align}
C_{s}^{\operatorname{Fk}}\approx\left\{\sum_{l=1}^{k}\log \frac{\Big|\rho\mathbf{h}_{s_l}^{\dagger}\mathbf{A}_{l-1}^{\bot} \mathbf{h}_{s_l} \Big|}{{\Big|\rho\mathbf{g}_{s_l}^{\dagger}\mathbf{B}_{l-1}^{\bot} \mathbf{g}_{s_l} \Big|}}\right\}^+.
\end{align}

From Algorithm 2, we know that the secrecy rate is maximized by choosing the $j$-th user with the largest $\frac{|\mathbf{h}_j^{\dagger}\mathbf{A}_{l-1}^{\bot}\mathbf{h}_j|}
{|\mathbf{g}_j^{\dagger}\mathbf{B}_{l-1}^{\bot}\mathbf{g}_j|}$ in each step. Therefore, the ESSR for the full CSI is given by
\begin{align}
C_{s}^{\operatorname{Fk\_gre}} &\approx \mathbb{E}\left\{ \sum_{l=1}^{k} \underset{1 \le j \le K-l+1}{\max} \log \frac{|\mathbf{h}_j^{\dagger}\mathbf{A}_{l-1}^{\bot}\mathbf{h}_j|}
{|\mathbf{g}_j^{\dagger}\mathbf{B}_{l-1}^{\bot}\mathbf{g}_j|}\right\}^+  \nonumber \\
&\gtrapprox \left\{ \sum_{l=1}^{k}\mathbb{E} \left[\underset{1 \le j \le K-l+1}{\max} \log \frac{|X_{lj}|}
{|Y_{lj}|}\right]\right\}^+.
\end{align}

Let $Z_{lj} = \log \frac{|X_{lj}|}
{|Y_{lj}|}$. With Lemma 1, $Z_{lj}$ can be also approximated as a Gaussian variable obeying $\mathcal{N}\big(\psi{(M_l)}-\psi{(N_l)},\sigma_{M_l}^2+\sigma_{N_l}^2\big)$, where $\sigma_{Nl}^2 = \frac{\pi^2}{6}-\sum_{i=1}^{N_l-1}\frac{1}{i^2}$. Then based on Lemma 2, we have
\begin{align}
\mathbb{E}\Big[{\underset{1 \le j \le K-l+1} {\max} Z_{lj}} \Big]= \psi{(M_l)}-\psi{(N_l)} + \sqrt{\sigma_{Ml}^2+\sigma_{Nl}^2}G_{K}^{l}. \nonumber
\end{align}

After some manipulations, the result in \eqref{Theorem_32} is obtained.

\begin{IEEEbiography}[{\includegraphics[width=1in,height=1.25in,clip,keepaspectratio]{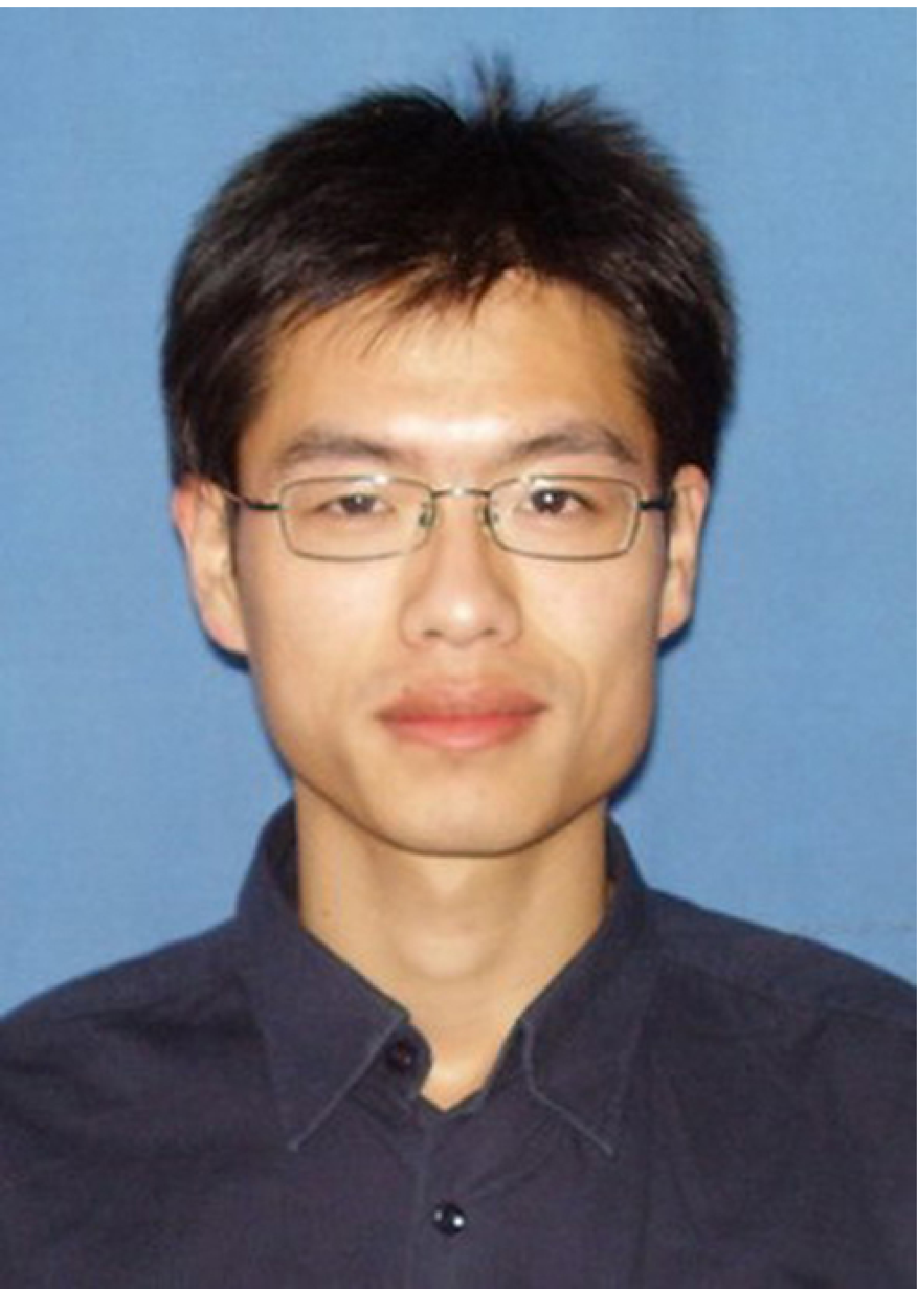}}]{Hao Deng}
is currently working towards the Ph.D at Xi'an Jiaotong University, and also a lecturer with the school of physics
and electronics, Henan University, Kaifeng, China. His research interests include cooperative communications and
physical layer security.
\end{IEEEbiography}

\begin{IEEEbiography}[{\includegraphics[width=1in,height=1.25in,clip,keepaspectratio]{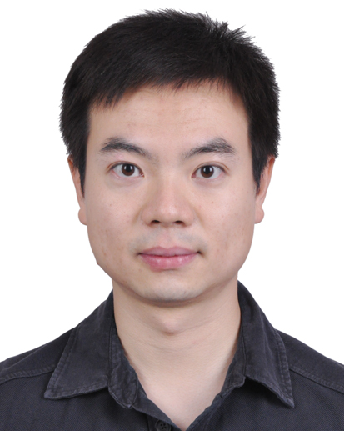}}]{Hui-Ming
Wang} (S'07--M'10--SM'16) received the B.S. and Ph.D. degrees, both with first class honors in Electrical Engineering from Xi’an Jiaotong University, Xi’an, China, in 2004 and 2010, respectively. He is currently a Full Professor with the Department of Information and Communications Engineering, Xi’an Jiaotong University, and also with the Ministry of Education Key Lab for Intelligent Networks and Network Security, China. From 2007 to 2008, and 2009 to 2010, he was a Visiting Scholar at the Department of Electrical and Computer Engineering, University of Delaware, USA. His research interests include cooperative communication systems, physical-layer security of wireless communications, MIMO and space-timing coding.

Dr. Wang received the National Excellent Doctoral Dissertation Award in China in 2012, a Best Paper Award of International Conference on Wireless Communications and Signal Processing, 2011, and a Best Paper Award of IEEE/CIC International Conference on Communications in China, 2014. He was a Symposium Chair of Wireless Communications and Networking in ChinaCom 2015. He has been a TPC member of various conferences, including the IEEE GlobeCom, ICC, WCNC, VTC, and PIMRC, etc. He was honored as an Exemplary Reviewer of the IEEE Transactions on Communications in 2016.
\end{IEEEbiography}

\begin{IEEEbiography}[{\includegraphics[width=1in,height=1.25in,clip,keepaspectratio]{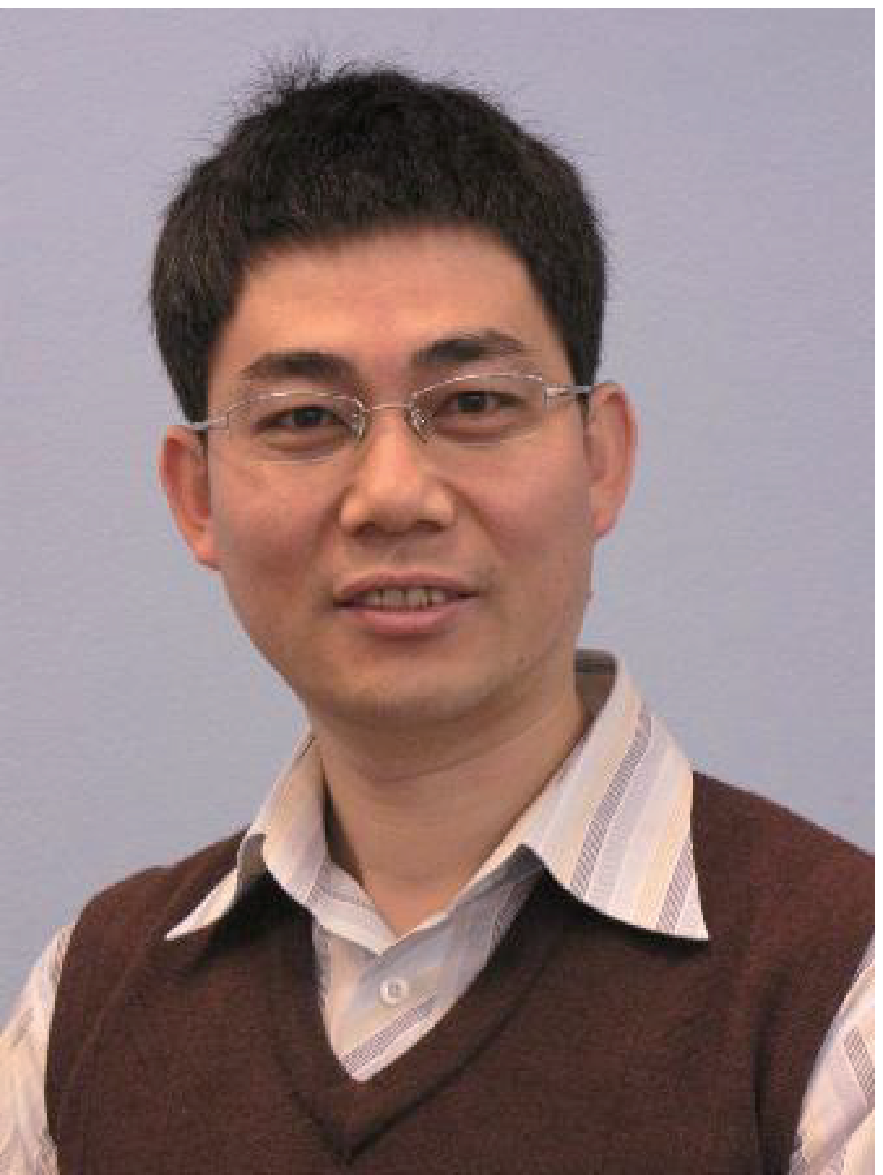}}]{Jinhong Yuan}
(M'02--SM'11--F'16) received the B.E. and Ph.D. degrees in electronics engineering from the Beijing Institute of Technology, Beijing, China, in 1991 and 1997, respectively. From 1997 to 1999, he was a Research Fellow with the School of Electrical Engineering, University of Sydney, Sydney, Australia. In 2000, he joined the School of Electrical Engineering and Telecommunications, University of New South Wales, Sydney, Australia, where he is currently a Telecommunications Professor with the School.

He has published two books, three book chapters, over 200 papers in telecommunications journals and conference proceedings, and 40 industrial reports. He is a co-inventor of one patent on MIMO systems and two patents on low-density-parity-check codes. He has co-authored three Best Paper Awards and one Best Poster Award, including the Best Paper Award from the IEEE Wireless Communications and Networking Conference, Cancun, Mexico, in 2011, and the Best Paper Award from the IEEE International Symposium on Wireless Communications Systems, Trondheim, Norway, in 2007. He is currently serving as an Associate Editor for the IEEE Transactions on Communications. He served as the IEEE NSW Chair of Joint Communications/Signal Processions/Ocean Engineering Chapter during 2011-2014. His current research interests include error control coding and information theory, communication theory, and wireless communications.
\end{IEEEbiography}

\begin{IEEEbiography}[{\includegraphics[width=1in,height=1.25in,clip,keepaspectratio]{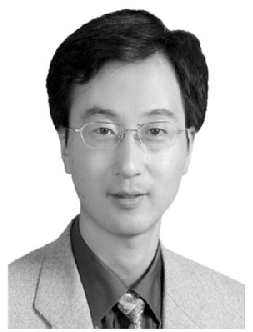}}]{Wenjie
Wang}
(M'10) received the B.S., M.S., and Ph.D. degrees in information and communication engineering from Xi’an Jiaotong
University, Xi’an,China, in 1993, 1998, and 2001, respectively.

From 2009 to 2010, he was a visiting scholar at the Department of Electrical and Computer Engineering, University of
Delaware, Newark. Currently, he is a Professor at Xi’an Jiaotong University. His main research interests include
MIMO and OFDM systems, digital signal processing, and wireless sensor networks.
\end{IEEEbiography}

\begin{IEEEbiography}[{\includegraphics[width=1in,height=1.25in,clip,keepaspectratio]{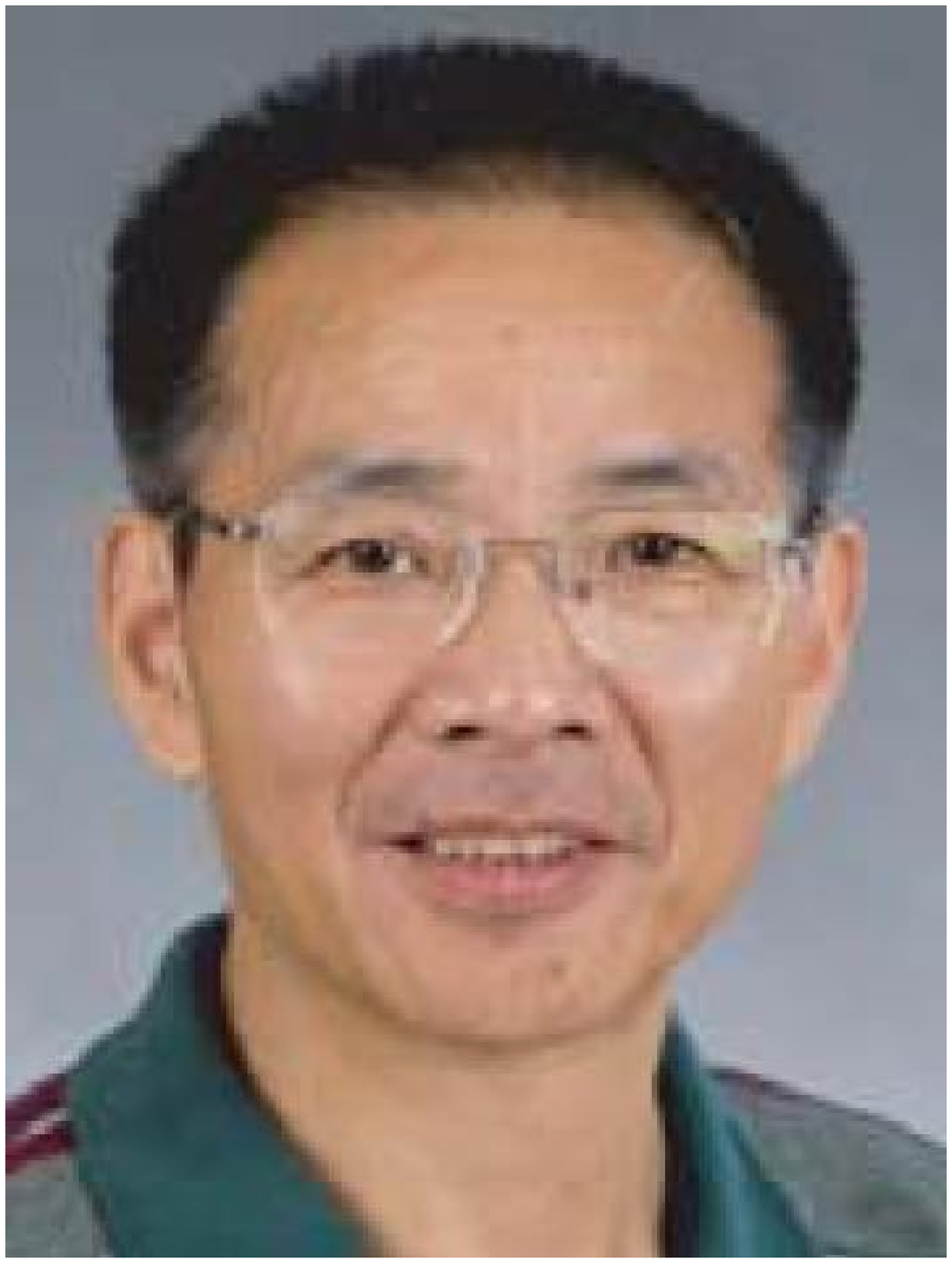}}]{Qinye
Yin}
received the B.S., M.S., and Ph.D. degrees in Communication and Electronic Systems from Xi'an Jiaotong University, Xi'an, China, in
1982, 1985, and 1989, respectively. Since 1989, he has been on the faculty at Xi'an Jiaotong University, where he is currently a
Professor of Information and Communications Engineering Department. From 1987 to 1989, he was a visiting scholar at the University of Maryland, MD, USA. From June to December 1996, he was a visiting scholar at the University of Taxes, Austin, TX, USA. His research interests focus on the joint time-frequency analysis and synthesis, multiple antenna MIMO broadband communication systems (including smart antenna systems), parameter estimation, and array signal processing.
\end{IEEEbiography}
\enlargethispage{-9.5cm}

\begin{thebibliography}{99}
\bibitem{ref:F. Oggier}
F. Oggier, and B. Hassibi, ``A perspective on the MIMO wiretap channel,'' \emph{Proceedings of the IEEE}, vol. 103, no. 10, pp. 1874-1882, oct. 2015.

\bibitem{ref:A. L.}
A. Mukherjee and A. L. Swindlehurst, ``Robust beamforming for security in MIMO wiretap channels with imperfect CSI,'' \emph{IEEE Trans. Signal Process.}, vol. 59, no. 1, pp. 351-361, Jan. 2011.

\bibitem{ref:Z. Rezki}
Z. Rezki and M.-S. Alouini, ``Secure diversity-multiplexing tradeoff zero-forcing transmit scheme at finite-SNR,'' \emph{IEEE Trans. Commun.}, vol.60, no. 4, pp. 1138-1147, Apr. 2012.

\bibitem{ref:Z. Ding}
Z. Ding, K. Leung, D. L. Goeckel and D. Towsley, ``On the application of cooperative transmission to secrecy Communications", \emph{IEEE J. Sel. Areas Commun.}, vol.30, no.2, pp.359-368, Feb. 2012.

\bibitem{ref:Wu2012TVT}
{Y. Wu, C. Xiao, Z. Ding, X. Gao, and S. Jin}, ``Linear precoding for finite alphabet signaling over MIMOME wiretap channels,'' \emph{IEEE Trans. Veh. Technol.}, vol.61, no.6, pp. 2599-2612, Jul. 2012.

\bibitem{ref:Wang Hybrid}
H.-M. Wang, Miao Luo, Q. Yin, and X.-G. Xia, ``Hybrid cooperative beamforming and jamming for physical-layer security of two-way relay networks,'' \emph{IEEE Trans. Inf. Foren. Sec.}, vol. 8, no. 12, pp. 2007-2020, Dec. 2013.

\bibitem{ref:T. Shang-Ho}
T. Shang-Ho and H. V. Poor, ``Power allocation for artificial-noise secure MIMO precoding systems,'' \emph{IEEE Trans. Signal Process.}, vol. 62, no. 13, pp. 3479-3493, Jul. 2014.

\bibitem{ref:Wu2016TIT}
{ Y. Wu, R. Schober, D. W. K. Ng, C. Xiao, and G. Caire}, ``Secure massive MIMO transmission with an active eavesdropper,'' \emph{IEEE Trans. Inf. Theory},  vol.62, no.7, pp. 3880-3900, Jul. 2016

\bibitem{ref:X. Zhou Achievablerate}
X. Zhou and M. R. McKay, ``Secure transmission with artificial noise over fading channels: achievable rate and optimal power allocation,'' \emph{IEEE Trans. Veh. Tech.}, vol. 59, no. 8, pp. 3831-3842, Oct. 2010.

\bibitem{ref:Wang TSP}
H.-M. Wang, C. Wang, and D. W. Kwan Ng, ``Artificial noise assisted secure transmission under training and feedback,'' \emph{IEEE Trans. Signal Process.}, vol. 63, no. 23, pp. 6285-6298, Dec. 2015.

\bibitem{ref:Wang TWC}
H.-M. Wang, T. Zheng, and X.-G. Xia, ``Secure MISO wiretap channels with multi-antenna passive eavesdropper: artificial noise vs. artificial fast fading,'' \emph{IEEE Trans. Wireless Commun.}, vol. 14, no. 1, pp. 94-106, Jan. 2015.

\bibitem{ref:A. Goldsmith}
T. Yoo and A. Goldsmith, ``On the optimality of multiantenna broadcast scheduling using zero-forcing beamforming,'' \emph{IEEE J.  Sel. Areas Commun.}, vol. 24, no. 3, pp. 528-541, Mar. 2006.

\bibitem{ref:M. Sharif Partial}
M. Sharif and B. Hassibi, ``On the capacity of MIMO broadcast channels with partial side information,'' \emph{IEEE Trans. Inf. Theory}, vol. 51, no. 2, pp. 506-522, Feb. 2005.

\bibitem{ref:R. Knopp}
R. Knopp and P. Humblet, ``Information capacity and power control in single-cell multiuser communications,'' \emph{in IEEE International Conference on Communications}, Seattle, vol. 1, 1995, pp. 331-335.

\bibitem{ref:G. Dimic}
G. Dimi\'c and N. D. Sidiropoulos, ``On downlink beamforming with greedy user selection: performance analysis and a simple new algorithm,'' \emph{IEEE Trans. Signal Process.}, vol. 53, no. 10, pp. 3857-3868, Oct. 2005.

\bibitem{ref:X. Qin}
X. Qin and R. A. Berry, ``Distributed approaches for exploiting multiuser diversity in wireless networks,'' \emph{IEEE Trans. Inf. Theory}, vol. 52, no. 2, pp. 392-413, Feb. 2006.

\bibitem{ref:J. Choi}
J. Choi and F. Adachi, ``User selection criteria for multiuser systems with optimal and suboptimal LR based detectors,'' \emph {IEEE Trans. Signal Process.}, vol. 58, no. 10, pp. 5463-5468, Oct. 2010.

\bibitem{ref:M. A. Maddah-Ali}
M. A. Maddah-Ali, M. Ansari, and A. K. Khandani, ``Broadcast in MIMO systems based on a generalized QR decomposition: Signaling and performance analysis,'' \emph{IEEE Trans. Inf. Theory}, vol. 54, no. 3, pp. 1124-1138, Mar. 2008.

\bibitem{ref:M. Pei CL}
M. Pei, A.L. Swindlehurst,  D. Ma, and J. Wei, ``On ergodic secrecy rate for MISO wiretap broadcast channels with opportunistic scheduling,'' \emph{IEEE Comm. Letter}, vol. 18, no. 1, pp. 50-53, Jan. 2014.

\bibitem{ref:X. Liu}
X. Liu, F. Gao, G. Wang, and X. Wang,  ``Joint beamforming and user selection in multicast downlink channel under secrecy-outage constraint,'' \emph{IEEE Comm. Letter}, vol. 18, no. 1, pp. 82-85, Jan. 2014.

\bibitem{T. M. Hoang}
 T. M. Hoang, T. Q.  Duong, H. A. Suraweera, C. Tellambura, and H. V. Poor, ``Cooperative beamforming and user selection for improving the security of relay-aided systems,'' \emph{IEEE Trans. Commun.}, vol. 63, no. 12, pp. 5039-5051, Dec. 12.

\bibitem{ref:J. Lee}
J. Lee and W. Choi,  ``Multiuser diversity for secrecy communications using opportunistic jammer selection: secure DoF and jammer scaling law,'' \emph{IEEE Trans. Signal Process.}, vol. 62, no. 4,  pp. 828-839, Aug. 2014.

\bibitem{ref:C Wang TWC}
C. Wang, H.-M. Wang, X.-G. Xia, and C. Liu, ``Uncoordinated jammer selection for securing SIMOME wiretap channels: A stochastic geometry approach,'' \emph{IEEE Trans. Wireless Commun.},  vol. 14, no. 5, pp. 2596-2612, May. 2015.

\bibitem{ref:C Wang SPL}
C. Wang, H.-M. Wang, and B. Wang, ``Low-overhead Distributed Jamming for SIMO Secrecy Transmission with Statistical CSI,'' \emph{IEEE Signal Process. Lett.}, vol. 22, no. 12, pp. 2294-2298, Dec. 2015.

\bibitem{ref:H. Alves}
H. Alves, R. D. Souza, M. Debbah, and M. Bennis,  ``Performance of transmit antenna selection physical layer security schemes,'' \emph{IEEE Signal Process. Lett.}, vol. 19, no. 6, pp. 372-375, Jun. 2012.

\bibitem{ref:N. Yang}
N. Yang, P. L. Yeoh, M. Elkashlan, R. Schober, and I. B. Collings,  ``Transmit antenna selection for security enhancement in MIMO wiretap channels,'' \emph{IEEE Trans. Commun.}, vol. 61, no. 1, pp. 144-154, Jan. 2013.

\bibitem{ref:Selection}
I. Krikidis, J. Thompson, and S. Mclaughlin, ``Relay selection for secure cooperative networks with jamming,'' \emph{IEEE Trans. Wireless Commun.},vol. 8, no. 10, pp. 5003-5011, Oct. 2009.

\bibitem{ref:A. Mabrouk}
A. Mabrouk, K. Tourki and N. Hamdi, ``Relay selection for optimized cooperative jamming scheme,'' \emph{ in 2015 23rd European Signal Processing Conference (EUSIPCO)}, Nice, 2015.

\bibitem{ref:Z. Wang}
Z. Wang and G. B. Giannakis, ``Outage mutual information of space-time MIMO channels," \emph{IEEE Trans. Inf. Theory}, vol. 50, no. 4, pp. 657-662, Apr. 2004.

\bibitem{ref:E. Biglieri}
E. Biglieri and G. Taricco, ``Transmission and reception with multiple antennas: theoretical foundations,'' \emph{Foundations and Trends in Communications and Information Theory}, vol. 1, no. 2, pp. 183-332, 2004.

\bibitem{ref:D. Tse}
D. Tse and P. Viswanath, \emph{Fundamentals of Wireless Communication}. Cambridge, U.K.: Cambridge Univ. Press, 2005.

\bibitem{ref:Duality Tse}
P. Viswanath and D. N. C. Tse, ``Sum capacity of the vector Gaussian broadcast channel and uplink-downlink duality,'' \emph{IEEE Trans. Inf. Theory}, vol. 49, no. 8, pp. 1912-1921, Aug. 2003.

\bibitem{ref:G. Bagherikaram}
G. Bagherikaram, A. S. Motahari, and A. K. Khandani, ``On the secure degrees of freedom of the multiple access channel,'' \emph{IEEE Trans. Inf. Theory}, 2010, submitted. [Online]. Available: arXrv:1003.0729.

\bibitem{ref:M. Pei}
M. Pei, A. L. Swindlehurst, D. Ma, and J. Wei, ``Adaptive limited feedback for MISO wiretap channels with cooperative jamming,'' \emph{IEEE Trans. Signal Process.}, vol. 62, no. 4, pp. 993-1004, Feb. 2014.

\bibitem{ref:Gopala}
P. Gopala, L. Lai, and H. El Gamal, ``On the secrecy capacity of fading channels,'' \emph{IEEE Trans. Inf. Theory}, vol. 54, no. 10, pp. 4687-4698, Oct. 2008.

\bibitem{ref:Y. Liang}
Y. Liang, H. Poor, and S. Shamai, ``Information theoretic security,'' \emph{Foundations and Trends in Communications
and Information Theory}, vol. 5, no. 4-5, pp. 355-580, 2009.

\bibitem{ref:B. M. Hochwald}
B. M. Hochwald, T. L. Marzetta, and V. Tarokh,  ``Multiple-antenna channel hardening and its implications for rate feedback and scheduling,'' \emph{IEEE Trans. Inf. Theory}, vol. 50, no. 9, pp. 1893-1909, Sep. 2004.

\bibitem{ref:Table}
I. S. Gradshteyn and I. M. Ryzhik, \emph{Table of Integrals, Series, and Products}, 7th ed. San Diego, CA: Academic, 1994.

\bibitem{ref:R. M. Young}
R. M. Young,  ``Euler’s constant,'' \emph{Math. Gaz.}, vol. 75, pp. 189-190, 1991.

\bibitem{J. Proakis}
J. Proakis and M. Salehi, \emph{Digital Communications}. McGraw-Hill Higher Education, 2008.

\bibitem{ref:D. Horgan}
D. Horgan and C. C. Murphy,  ``On the convergence of the Chi square and noncentral chi square distributions to the normal distribution,'' \emph{IEEE Comm. Letter}, vol. 17, no. 12, pp. 2233-2236, Dec. 2013.

\bibitem{ref:H. Q. Ngo}
H. Q. Ngo, E. G. Larsson, and T. L. Marzetta, ``Energy and spectral efficiency of very large multiuser MIMO systems,'' \emph{IEEE Trans. Commun.}, vol. 61, pp. 1436-1449, Apr. 2013.

\bibitem{ref:Meyer}
C. D. Meyer, \emph{Matrix analysis and applied linear algebra}, Siam, 2000.

\bibitem{ref:D. Bai}
D. Bai, P. Mitran, S. Ghassemzadeh, R. R. Miller, and V. Tarokh, ''Rate of channel hardening of antenna selection diversity schemes and its implication on scheduling," \emph{IEEE Trans. Inf. Theory}, vol. 55, pp. 4353-4365, Oct. 2009.

\bibitem{ref:J. Kampeas}
J. Kampeas, A. Cohen, and O. Gurewitz,  ``The capacity of the multiple access channel under distributed scheduling - order optimality of linear receivers,'' \emph{Submitted to IEEE Trans. Inf. Theory (arXiv preprint 1304.7480)}, 2013.

\bibitem{ref:David}
H. A. David and H. N. Nagaraja, \emph{Order Statistics}, 3rd ed., New Jeresy: John Wiley \& Sons, 2003.

\bibitem{ref:Bletsas}
A. Bletsas, A. Khisti, D. P. Reed, and A. Lippman, ``A simple cooperative diversity method based on network path selection,''\emph{ IEEE Journal on Select. Areas in Comm.}, vol. 24, pp. 659-672, Mar. 2006.

\bibitem{ref:Yoo}
T. Yoo and A. Goldsmith, ``Capacity and power allocation for fading MIMO channels with channel estimation error," \emph{IEEE Trans. Inf. Theory}, vol. 52, pp. 2203-2214, May 2006.

\bibitem{ref:R. L. Smith}
R. L. Smith,  ``Extreme value theory based on the $r$ largest annual events,'' \emph{Journal of Hydrology}, pp. 27-43, 1986.

\bibitem{ref:Mathew}
T. Mathew and K. Nordstr{\"{o}}m, ``Wishart and Chi-square distributions associated
with matrix quadratic forms," \emph{J. Multivariate Analysis}, vol. 61, pp. 129-143, 1997.

\bibitem{ref:S. Ozyurt}
S. Ozyurt and M. Torlak,  ``Exact joint distribution analysis of zero-forcing V-BLAST gains with greedy ordering,'' \emph{IEEE Trans. Commun.}, vol. 12, no. 11, pp. 5377-5385, Nov. 2013.

%\vspace{-1cm}
\end{thebibliography}
\end{document}